\documentclass[a4paper,11pt]{article}
\pdfoutput=1 

\usepackage{jcappub} 

\usepackage[british]{babel}
\usepackage{newtxtext}
\usepackage[slantedGreek]{newtxmath}
\usepackage[T1]{fontenc} 
\usepackage{graphicx}	
\usepackage{gensymb}
\usepackage{amsmath}	
\usepackage{amssymb}	
\usepackage{multirow}
\usepackage{pbox}
\usepackage{tabularx}
\usepackage{easy-todo}	
\usepackage{pifont}
\usepackage[normalem]{ulem}
\usepackage{placeins}
\usepackage[inline]{enumitem}
\usepackage{soul} 
\usepackage{amsmath}
\usepackage{xifthen}
\usepackage{xparse}
\usepackage{xspace}

\newcommand{\orcid}[1]{\href{https://orcid.org/#1}{\,\includegraphics[height=\fontcharht\font`\B]{ORCIDiD.pdf}}}
\RenewDocumentCommand{\citet}{
  O{} 
  O{} 
  m
  }{%
  \citep[#2][#1]{#3}%
}

\newcommand{\WMAPsevenp}{\ensuremath{{\Ho{}=\kms[70.4]\pMpc{}},\,\allowbreak {\Omegaup_{\rm M}=0.272},\,\allowbreak {\Omegaup_\Lambdaup=0.728},\,\allowbreak {n_s=0.967},\, \sigma_8=0.81}}

\newcommand{\fviable}{\ensuremath{f_{\rm v}}}

\newcommand{\dv}[1]{\ensuremath{\mathrm{d}#1}} 
\newcommand{\fdv}[2]{\ensuremath{\frac{\mathrm{d}#1}{\mathrm{d}#2}}} 

\newcolumntype{L}{>{\raggedright\arraybackslash}X}
\newcolumntype{Y}{>{\centering\arraybackslash}X}
\newcolumntype{R}{>{\raggedleft\arraybackslash}X}

\newcommand{\appref}[1]{Appendix~\ref{#1}}
\newcommand{\secref}[1]{Section~\ref{#1}}
\newcommand{\secrefs}[2]{Sections~\ref{#1} and \ref{#2}}
\newcommand{\figref}[1]{Figure~\ref{#1}}
\newcommand{\figrefs}[2]{Figures~\ref{#1}~and~\ref{#2}}
\newcommand{\eqnref}[1]{equation~(\ref{#1})}

\newcommand{\extapp}[1]{appendix~#1}

\newcommand{\extfig}[1]{fig.~\ensuremath{#1}}
\newcommand{\extsec}[1]{section~\ensuremath{#1}}

\newcommand{\Myref}[1]{Ref.~\citep{#1}}
\newcommand{\myref}[1]{ref.~\citep{#1}}
\newcommand{\myrefs}[2]{refs.~\citep{#1} and \citep{#2}}
\newcommand{\tabref}[1]{Table~\ref{#1}}

\defcitealias{newton_total_2018}{N18}
\defcitealias{lacey_unified_2016}{L16}
\defcitealias{parkinson_generating_2008}{PCH}
\newcommand{\lacey}{\citetalias{lacey_unified_2016}}

\newcommand{\MConstraint}[1]{\ensuremath{\Mth{\leq}\keV[#1]}}
\newcommand{\MthTopConstraint}{\MConstraint{1.80}} 
\newcommand{\MthGFTopConstraint}{\MConstraint{2.95}} 
\newcommand{\marginalisedDMConstraint}{\MConstraint{2.02}} 
\newcommand{\marginalisedGalformConstraint}{\MConstraint{\zsevenvthree{}}} 

\newcommand{\zsixvtwofive}{2.86}
\newcommand{\zsixvthree}{3.37}
\newcommand{\zsixvthreefive}{3.52}
\newcommand{\zsevenvtwofive}{3.12}
\newcommand{\zsevenvthree}{3.99}
\newcommand{\zsevenvthreefive}{4.37}
\newcommand{\zeightvtwofive}{3.49}
\newcommand{\zeightvthree}{5.26}
\newcommand{\zeightvthreefive}{5.82}

\newcommand{\newtonheadline}{\ensuremath{124^{+40}_{-27}}}   

\newcommand{\Edges}{{\sc EDGES}}

\newcommand{\Aquarius}{Aquarius}
\newcommand{\COCO}{\mbox{{\sc COCO}}}
\newcommand{\cCOCO}{\mbox{{\sc COCO-COLD}}}
\newcommand{\wCOCO}{\mbox{{\sc COCO-WARM}}}

\newcommand{\Galform}{{\sc galform}}

\newcommand{\Gadget}{{\sc gadget}}

\newcommand{\GadgetIII}{\Gadget3}

\newcommand{\astropy}{{\sc Astropy}}
\newcommand{\matplotlib}{{\sc matplotlib}}
\newcommand{\numpy}{{\sc numpy}}
\newcommand{\python}{{\sc python}}
\newcommand{\scipy}{{\sc scipy}}
\newcommand{\Subfind}{{\sc subfind}}

\newcommand{\CDM}{{CDM}}

\newcommand{\hSpher}{\ensuremath{s_\text{half-max}}}

\newcommand{\LCDM}{{$\Lambdaup$CDM}}

\newcommand{\Lyman}[1]{Ly~\ensuremath{#1}}

\newcommand{\Mth}{\ensuremath{m_{\rm th}}}

\newcommand{\Nbody}{\mbox{\textit{N}--body}}
\newcommand{\Nsat}[1][]{%
  \ifthenelse{\isempty{#1}}%
    {\ensuremath{N_{\rm sat}}}
    {\ensuremath{N^{#1}_{\rm sat}}}
  }
\newcommand{\Nsub}[1][]{%
  \ifthenelse{\isempty{#1}}%
    {\ensuremath{N_{\rm sub}}}
    {\ensuremath{N^{#1}_{\rm sub}}}
  }
\newcommand{\R}[1]{\ensuremath{R_{#1}}}
\newcommand{\RNFW}{\R{200}}
\newcommand{\vcut}[1][]{%
  \ifthenelse{\isempty{#1}}%
    {\ensuremath{V_{\rm cut}}}
    {\ensuremath{V_{\rm cut}{=}\kms[{#1}]}}
  }

\newcommand{\WDM}{{WDM}}

\newcommand{\DES}{{DES}}

\newcommand{\LSST}{{LSST}}

\newcommand{\PAN}{{Pan-STARRS}}

\newcommand{\SDSS}{{SDSS}}

\newcommand{\WMAP}{{WMAP}}

\newcommand{\unit}[1]{\ensuremath{\mathrm{\,#1}}\xspace}
\newcommand{\unitlogicnospace}[2]{%
  \ifthenelse{\isempty{#1}}%
    {\unit{#2}}
    {\ensuremath{{#1}\unit{#2}}}
  }
\newcommand{\unitlogicspace}[2]{%
  \ifthenelse{\isempty{#1}}%
    {\unit{#2}}
    {\ensuremath{{#1}\, \unit{#2}}}
  }

\newcommand{\cMpc}[1][]{%
  \ifthenelse{\isempty{#1}}%
    {h^{-1}\! \Mpc{}}
    {\ensuremath{{#1}\, h^{-1}\! \Mpc{}}}
  }
\newcommand{\cMsun}[1][]{%
  \ifthenelse{\isempty{#1}}%
    {h^{-1}\! \Msun{}}
    {\ensuremath{{#1}\, h^{-1}\! \Msun{}}}
  }
\newcommand{\cm}[1][]{\unitlogicspace{#1}{cm}}
\newcommand{\keV}[1][]{\unitlogicspace{#1}{keV}}
\newcommand{\kms}[1][]{\unitlogicspace{#1}{km\, s^{-1}}}
\newcommand{\kpc}[1][]{\unitlogicspace{#1}{kpc}}

\newcommand{\Mpc}[1][]{\unitlogicspace{#1}{Mpc}}
\newcommand{\Msun}[1][]{\unitlogicspace{#1}{M_\odot}}
\newcommand{\pMpc}[1][]{\unitlogicspace{#1}{Mpc^{-1}}}

\newcommand{\percent}[1]{\ensuremath{#1}~per cent}

\newcommand{\variablelogicspace}[2]{%
  \ifthenelse{\isempty{#2}}%
    {\ensuremath{#1}}
    {\ensuremath{{#1}{=}{#2}}}
  }

\newcommand{\Ho}{\ensuremath{H_0}}

\newcommand{\vpeak}{\ensuremath{v_\mathrm{peak}}}
\newcommand{\Mpeak}{\ensuremath{M_\mathrm{peak}}}
\newcommand{\MNFW}[1][]{\variablelogicspace{M_{200}}{#1}}
\newcommand{\MV}[1][]{\variablelogicspace{M_\mathrm{V}}{#1}}
\newcommand{\z}[1][]{\variablelogicspace{z}{#1}}
\newcommand{\zreion}[1][]{\variablelogicspace{z_\mathrm{reion}}{#1}}
\graphicspath{{Images/}}

\title{Constraints on the properties of warm dark matter using the satellite galaxies of the Milky Way}

\author[a,b]{Oliver~Newton,}
\author[1]{Matteo~Leo,\note{Independent researcher, Bari, 70125, Italy}}
\author[a,c]{Marius~Cautun,}
\author[a]{Adrian~Jenkins,}
\author[a]{Carlos~S.~Frenk,}
\author[a,d]{Mark~R.~Lovell,}
\author[a]{John~C.~Helly}
\author[e]{Andrew~J.~Benson}
\author[a]{Shaun~Cole}

\affiliation[a]{Institute for Computational Cosmology, Durham University, South Road, Durham, UK}
\affiliation[b]{Univ. Lyon, UCBL~1, CNRS, IP2I Lyon/IN2P3, IMR 5822, F-69622 Villeurbanne, France}
\affiliation[c]{Leiden Observatory, Leiden University, NL-2300 RA Leiden, the Netherlands}
\affiliation[d]{Center for Astrophysics and Cosmology, Science Institute, University of Iceland\\107 Reykjav\'{i}k, Iceland}
\affiliation[e]{Carnegie Observatories, 813 Santa Barbara Street, Pasadena, CA 91101, U.S.A}

\emailAdd{olivier.newton@univ-lyon1.fr}
\emailAdd{cautun@strw.leidenuniv.nl}
\emailAdd{a.r.jenkins@durham.ac.uk}
\emailAdd{c.s.frenk@durham.ac.uk}
\emailAdd{lovell@hi.is}
\emailAdd{j.c.helly@durham.ac.uk}
\emailAdd{abenson@carnegiescience.edu}
\emailAdd{shaun.cole@durham.ac.uk}

\abstract{%
The satellite galaxies of the Milky Way~(MW) are effective probes of the underlying dark matter~(DM) substructure, which is sensitive to the nature of the DM particle. In particular, a class of DM models have a power spectrum cut-off on the mass scale of dwarf galaxies and thus predict only small numbers of substructures below the cut-off mass. This makes the MW satellite system appealing to constrain the DM properties: feasible models must produce enough substructure to host the number of observed Galactic satellites. Here, we compare theoretical predictions of the abundance of DM substructure in thermal relic warm DM~(\WDM{}) models with estimates of the total satellite population of the MW. This produces conservative robust lower limits on the allowed mass, \Mth{}, of the thermal relic \WDM{} particle. As the abundance of satellite galaxies depends on the MW halo mass, we marginalize over the corresponding uncertainties and rule out \marginalisedDMConstraint{} at \percent{95} confidence \emph{independently} of assumptions about galaxy formation processes. Modelling some of these --- in particular, the effect of reionization, which suppresses the formation of dwarf galaxies --- strengthens our constraints on the DM properties and excludes models with \marginalisedGalformConstraint{} in our fiducial model. We also find that thermal relic models cannot produce enough satellites if the MW halo mass is $M_{200}\leq 0.6\times \Msun[10^{12}]$, which imposes a lower limit on the MW halo mass in \CDM{}. We address several observational and theoretical uncertainties and discuss how improvements in these will strengthen the DM mass constraints.
}%


\arxivnumber{2011.08865}

\begin{document}
\maketitle
\flushbottom



\section{Introduction}
\label{sec:Introduction}
Recent astrophysical observations have provided tentative indirect evidence for a candidate dark matter~(DM) particle with mass in the \keV{} range, e.g. \citep{boyarsky_unidentified_2014, bulbul_detection_2014}. Such a particle would be incompatible with the mass range proposed for candidate cold DM~(\CDM{}) particles and could have very different clustering properties on small scales \citep{boyarsky_unidentified_2014,boyarsky_checking_2015,bulbul_detection_2014,cappelluti_searching_2018}. This, together with a lack of any experimental detection of a \CDM{} particle despite considerable advances in particle detector technology \citep{liu_current_2017,xenon_collaboration_first_2017}, has motivated a renewed interest in possible alternatives to \CDM{}  \citep{boehm_using_2014,marsh_axion_2016,escudero_fresh_2018}. These seek to replicate the success of \CDM{} on large scales and to explain the observed small-scale features of \LCDM{} \citep{sawala_apostle_2016} with less reliance on `baryonic processes'. One family of these alternative DM models posits a `warm' DM~(\WDM{}) particle that would have a much higher thermal velocity than its \CDM{} counterpart at early times in the evolution of the Universe. These `thermal relics' are formed in equilibrium with the primordial plasma with masses such that they are relativistic at decoupling but non-relativistic by matter-radiation equality \citep{avila-reese_formation_2001,bode_halo_2001}. Such particles would free-stream out of small-scale primordial density perturbations, preventing their condensation into small haloes and producing a cut-off in the linear matter power spectrum on astrophysically relevant scales. Detecting this suppression of structure relative to \CDM{} predictions would provide a means of discriminating between the prevailing cosmological paradigm and viable \WDM{} models. The goal of this paper is to use visible tracers of the DM substructure to rule out thermal relic \WDM{} models that do not produce enough subhaloes to host the observed number of low-mass satellites of the Milky Way~(MW).

Low mass, DM-dominated galaxies provide an excellent probe of the `small-scale' DM structure \citep{shen_baryon_2014,sawala_bent_2015,sawala_chosen_2016,wheeler_sweating_2015}. The smallest and faintest of these can be observed best in the environs of the MW; however, the current census of ${\sim}60$ satellite galaxies is highly incomplete as extant surveys do not cover the entire sky to sufficient depth and large parts of it are partially or totally obscured by the MW itself \citep{koposov_luminosity_2008,walsh_invisibles_2009,hargis_too_2014}. Simple volume corrections to the observed complement of satellite galaxies have been used already to constrain the viable parameter space of thermal relic \WDM{} models by comparing the number of DM substructures in MW-mass haloes with the number of observed satellites \citep{kennedy_constraining_2014,lovell_properties_2014}. Such approaches make assumptions about the completeness of the surveys, which could lead to a misestimation of the real satellite population. More recent estimates of the satellite galaxy luminosity function that account for the stochasticity of observational data and uncertainties arising from the variability of host haloes at fixed halo mass suggest that the size of the total complement of MW satellites could be several times larger than previously assumed \citep{hargis_too_2014,newton_total_2018,nadler_modeling_2019}.

This paper improves on previous work and strengthens the methodology used to constrain the properties of candidate \WDM{} particles in several important ways, which we demonstrate using the thermal relic class of \WDM{} models. First, we use one of the most recent estimates of the total satellite population of the MW, which takes advantage of recent observational data to infer a population of \newtonheadline{} satellites brighter than \MV[0] within \kpc[300] of the Sun \citep{newton_total_2018}. This properly accounts for the incompleteness of current surveys; the method used to obtain this estimate has been tested robustly using mock observations. Secondly, our results account for resolution effects in \Nbody{} simulations that prevent the identification of DM subhaloes that survive to the present day but fall below the resolution limit of subhalo finders or are destroyed by numerical effects that enhance tidal stripping \citep{springel_aquarius_2008,onions_subhaloes_2012,van_den_bosch_dark_2018}. These significant effects have been overlooked in previous studies which, as a result, produce constraints on the viable parameter space of \WDM{} models that are too restrictive. Finally, we incorporate the uncertainty in the total number of satellite galaxies, which has not been included in previous analyses.

We organize this paper as follows. In \secref{sec:Methods}, we describe the method to constrain the properties of \WDM{} models by comparing their predictions of the abundance of subhaloes in MW-mass haloes with estimates of the total number of MW satellite galaxies from observations. We apply this methodology to thermal relic \WDM{} and present our main results in \secref{sec:Constraints_on_mth}. We investigate further the effect of reionization on the constraints that we obtain in \secref{sec:Galaxy_formation}. In \secref{sec:Discussion}, we discuss the implications of our results and consider some of the limitations of our method; we present concluding remarks in \secref{sec:Conclusions}.

\section{Methodology}
\label{sec:Methods}
Our goal is to use the satellite luminosity function of our Galaxy to constrain the properties of \WDM{} models using a minimal set of assumptions. In DM cosmologies, galaxies of all masses form almost exclusively within DM haloes.\footnote{Dwarf galaxies can also form during the collision of gas-rich massive galaxies and these are known as `tidal dwarf galaxies', e.g.~\citep{kaviraj_tidal_2012,lisenfeld_molecular_2016,ploeckinger_tidal_2018,haslbauer_galaxies_2019}. These are low mass and possess negligible DM content; consequently, they are thought to be short-lived. As our Galaxy has not experienced any recent major mergers, the MW is unlikely to contain a significant population of tidal dwarf galaxies.} The abundance of these can be probed readily with numerical simulations \citep{frenk_formation_1988} which provide a useful tool to investigate the predictions of different models; we introduce these in \secref{sec:Methods:Simulations}. A DM model is viable only if it forms enough subhaloes to host each MW satellite galaxy. To test for this condition we need two ingredients. First, we need an accurate estimate of the MW satellite galaxy luminosity function, which we discuss in \secref{sec:Methods:Model-independent_LF}. Secondly, we need a model to predict the number of substructures given the properties of the \WDM{} particle and the mass of the host DM halo, which we describe in \secref{sec:Methods:EPS}.

\subsection{\Nbody{} simulations}
\label{sec:Methods:Simulations}

We calibrate our predictions for the number of substructures in a WDM model using high-resolution DM-only \Nbody{} simulations of cosmological volumes. The Copernicus Complexio~(\COCO{}) suite consists of two zoom-in simulations: one of \LCDM{} that we refer to as \cCOCO{} \citep{hellwing_copernicus_2016}, and the other of \keV[3.3] thermal relic \WDM{}, hereafter \wCOCO{} \citep{bose_copernicus_2016}. These two versions differ only in the matter power spectra used to perturb the simulation particles in the initial conditions. Both \cCOCO{} and \wCOCO{} are simulated in periodic cubes of side \cMpc[70.4] using the \GadgetIII{} code that was developed for the \Aquarius{} Project \citep{springel_aquarius_2008}. The high-resolution regions correspond approximately to spherical volumes of radii \cMpc[{\sim}18] that each contain ${\sim}1.3\times10^{10}$ DM particles of mass, $m_p{=}\cMsun[1.135\times10^5]{}$. Haloes at the edges of these regions can become contaminated with high-mass simulation particles that disrupt their evolution. We identify these contaminated haloes as having a low-resolution DM particle within $3\times\RNFW{}$ of the halo centre at \z[0]. The cleaned catalogues provide large samples of haloes in both cosmological models and both simulations resolve the subhalo mass functions of DM haloes down to masses $\Msun[{\sim}10^7]$. The cosmological parameters assumed for this suite of simulations are derived from the \WMAP{} seventh-year data release \citep{komatsu_seven-year_2011}: \WMAPsevenp{}.

In \Nbody{} cosmological simulations the discreteness of the simulation particles can give rise to gravitational instabilities that produce artificial structures. Models such as \WDM{} that impose a cut-off in the primordial matter power spectrum are especially susceptible to these effects \citep{wang_discreteness_2007,angulo_warm_2013,lovell_properties_2014}. The instabilities are resolution-dependent and lead to the artificial fragmentation of filaments, giving rise to small `spurious' haloes that create an upturn at the low-mass end of the \WDM{} halo mass function. \Myref{lovell_properties_2014} developed a method to identify and prune these objects from the halo merger trees using their mass and particle content. The onset of numerical gravitational instabilities translates into a resolution-dependent mass threshold. Haloes that do not exceed this during their formation and subsequent evolution are likely to be spurious. This coarse requirement is refined further by a second criterion on the particles that compose the halo when its mass is half that of its maximum value, $M_{\rm max}\, /\, 2$. In the initial conditions of the simulation, the Lagrangian regions formed by the particles in spurious haloes are highly aspherical. Their shapes are  parametrized by $\hSpher{}{=}c\, /\, a$, where \emph{a} and \emph{c} are the major and minor axes of the diagonalized moment of inertia tensor of the DM particles in Lagrangian coordinates. These considerations were applied to the \wCOCO{} simulation by \myref{bose_copernicus_2016} who find that
almost all spurious haloes can be removed by applying the criteria: $M_{\rm max}{<}\cMsun[3.1\times10^7]$ and $\hSpher{}{<}0.165$.
The details of the calculation of these threshold values can be found in \extsec{2.3} of \myref{bose_copernicus_2016}.
Applying such simple criteria means that some genuine haloes can be removed while some spurious haloes remain; however, the numbers of each are extremely small and do not affect our results.
Therefore, we follow this prescription to `clean' the \wCOCO{} catalogues of spurious haloes for use throughout the rest of this paper.

The resolution of a simulation also affects the identification of subhaloes in the inner regions of simulated haloes, e.g. \citep{springel_aquarius_2008,onions_subhaloes_2012}. Subhaloes that fall below the resolution limit at any time are discarded by some substructure finders, and some other subhaloes are disrupted artificially by numerical effects \citep{van_den_bosch_dark_2018,green_tidal_2019,errani_asymptotic_2021,green_tidal_2021}. Consequently, these objects do not appear in the subhalo catalogue, even though they may still exist at the present day. We correct for this by identifying such subhaloes in \cCOCO{} and \wCOCO{} before they are accreted and tracking them to \z[0], and restoring them to the subhalo catalogues that we use to calibrate our methodology.
In \appref{app:Convergence_tests}, we discuss in more detail the procedure we use to recover these objects, and the effect that excluding them has on the halo mass function.

\subsection{Model-independent radial density profile of the MW satellites}
\label{sec:Methods:Model-independent_LF}

\begin{figure}%
    \centering%
	\includegraphics[width=0.5\columnwidth]{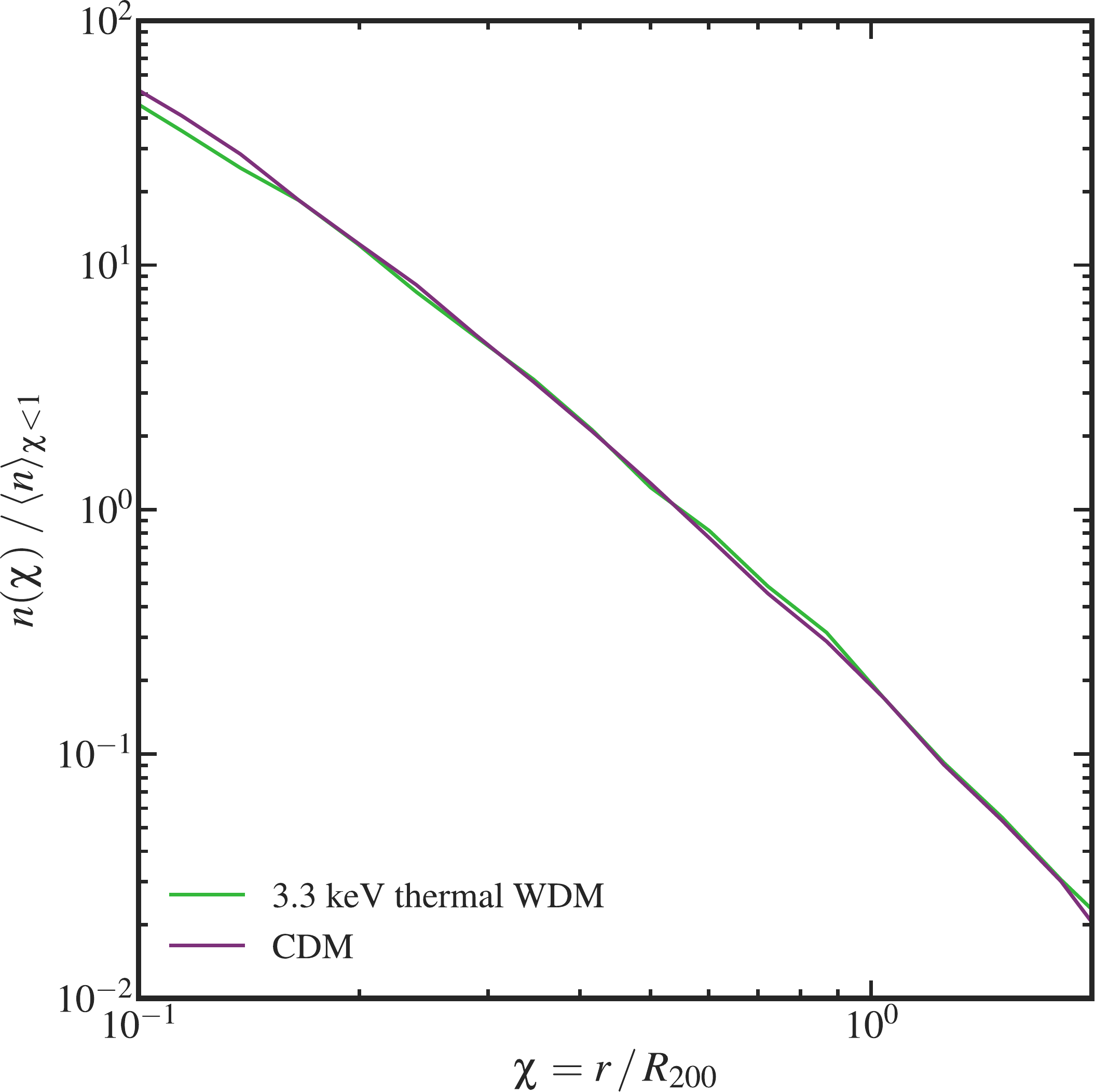}%
	\vspace{-10pt}%
	\caption[Stacked subhalo radial number density profiles of \COCO{} haloes]{The radial number density of subhaloes with ${\vpeak\geq\kms[20]}$ normalized to the mean density within \R{200}. The solid lines show the profiles obtained by stacking $767$ and $764$ uncontaminated host haloes with masses $\MNFW{}\geq\Msun[10^{11}]$ from \cCOCO{} and \wCOCO{}, respectively. The \percent{68} bootstrapped uncertainties in the stacked profiles are approximately the same size as the line thicknesses and are not shown.}
	\label{fig:Methods:compare_COCO_C_vs_W}%
	\vspace{-10pt}%
\end{figure}%

To obtain the best constraints on the \WDM{} particle mass we need a complete census of the Galactic satellites. The satellite population is dominated by ultra- and hyperfaint galaxies with absolute magnitudes fainter than $\MV{} = -8$ (see e.g. \citep{tollerud_hundreds_2008,hargis_too_2014,newton_total_2018}), which can be detected only in deep surveys. This means that large areas of the sky remain unexplored and that currently we have only a partial census of the MW satellites. However, there are several methods that use the current observations to infer the total satellite count of our Galaxy (see e.g. \citep{koposov_luminosity_2008,tollerud_hundreds_2008}). Here, we use the estimates from \myref{newton_total_2018} that are based on a Bayesian formalism that has been tested robustly using mock observations. These results were obtained by combining the observations of the Sloan Digital Sky Survey~(\SDSS{}) \citep{alam_eleventh_2015} and the Dark Energy Survey~(\DES{}) \citep{bechtol_eight_2015,drlica-wagner_eight_2015}, which together cover nearly half the sky, and estimating the MW satellite luminosity function down to a magnitude, \MV[0]. This roughly corresponds to galaxies with stellar mass higher than $\Msun[10^2]$ \citep{bose_no_2019}.

The method of \myref{newton_total_2018}~(code implementing this is available from \citep{newton_mw_2018}) takes two input components. First, it uses the sky coverage of a given survey and the distance from the Sun within which a satellite galaxy of a given magnitude can be detected. This depends on the depth of the survey and the satellite detection algorithm. Secondly, the \myref{newton_total_2018}  method requires the radial probability distribution function of satellite galaxies. Simulations of DM-only \CDM{} haloes show that subhaloes selected by \vpeak{}, the highest maximum circular velocity achieved in their evolutionary histories, have the same radial number density profile as that of the observed satellites (see \myref{newton_total_2018}, and discussion therein). Furthermore, \CDM{} simulations (e.g. \citep{springel_aquarius_2008,hellwing_copernicus_2016}) have shown that the radial distribution of satellites is largely independent of their mass as well as of the host mass when expressed in terms of the rescaled distance, $r/R_{200}$, where $r$ and $R_{200}$ denote the radial distance and the host halo radius, respectively. This is studied further in \figref{fig:Methods:compare_COCO_C_vs_W}, where we compare the normalized radial number density profiles of stacked populations of subhaloes in the \wCOCO{} and \cCOCO{} simulations. The fiducial populations were obtained by selecting subhaloes with ${\vpeak{}\geq\kms[20]}$ and identifying and including subhaloes that would exist at $z{=}0$ if they had not been prematurely destroyed or missed by substructure finders (for details see \appref{app:Convergence_tests}). We apply this correction after pruning the spurious haloes from the merger trees (see \secref{sec:Methods:Simulations}) to ensure that they are not inadvertently restored. \figref{fig:Methods:compare_COCO_C_vs_W} illustrates that both \CDM{} and \WDM{} predict the same radial distribution of satellites, which means that we can use the satellite distribution inferred from \CDM{} to make predictions for \WDM{} models. This is beneficial as \CDM{} simulations sample better the inner radial profile, to which the \myref{newton_total_2018} result is particularly sensitive.

To summarize, in this paper we infer the satellite galaxy luminosity function of the MW within \RNFW{} for assumed host halo masses in the range, $\MNFW{}=\Msun[{\left[0.5,\, 2.0\right]\times10^{12}}]$, using the Bayesian methodology presented in \myref{newton_total_2018}. As we mentioned above, this requires two components:
\begin{enumerate}
    \item a tracer population of DM subhaloes with a radial profile that matches that of the observed satellites; and,
    \item a set of satellite galaxies detected in surveys for which the completeness is characterized well.
\end{enumerate}
For the former, we use the same \vpeak{}-selected $\left(\vpeak{}{\geq}\kms[10]\right)$ fiducial \CDM{} subhalo populations as used in \myref{newton_total_2018}. These are obtained from five high-resolution \LCDM{} DM-only \Nbody{} simulations of isolated MW-like host haloes from the \Aquarius{} suite of simulations \citep{springel_aquarius_2008}. For the latter, we use the observations of nearby dwarf galaxies from the \SDSS{} and \DES{} supplied in \extapp{A} of \myref{newton_total_2018} (compiled from \citep{watkins_substructure_2009,mcconnachie_observed_2012,drlica-wagner_eight_2015,kim_heros_2015,koposov_kinematics_2015,jethwa_magellanic_2016,kim_portrait_2016,walker_magellan/m2fs_2016,carlin_deep_2017,li_farthest_2017}). Later work to infer the luminosity function using more recent observational data and a better characterization of the \DES{} completeness function is in good agreement with the \myref{newton_total_2018} results \citep{nadler_modeling_2019,drlica-wagner_milky_2020}.

\subsection{Estimating the amount of halo substructure}
\label{sec:Methods:EPS}
Estimates of the average number of subhaloes in MW-like DM haloes can be obtained using the Extended Press-Schechter~(EPS) formalism \citep{press_formation_1974,bond_excursion_1991,bower_evolution_1991,lacey_merger_1993,parkinson_generating_2008}. In this approach, the linear matter density field is filtered with a window function to identify regions that are sufficiently dense to collapse to form virialized DM haloes. In \CDM{} models the filter employed takes the form of a top-hat in real space. However, applying this to models such as \WDM{} in which power is suppressed at small scales leads to an over-prediction of the number of low-mass haloes \citep{benson_dark_2013}. This occurs because the variance of the smoothed density field on small scales becomes independent of the shape of the linear matter power spectrum if the latter decreases faster than $k^{-3}$. Consequently, the halo mass function continues to increase at small masses rather than turning over [\citealp{lovell_satellite_2016}, \citealp{leo_new_2018} \extsec{3.1}], making the top-hat filter an inappropriate choice. Using a sharp \emph{k}-space filter seemed to address this by accounting for the shape of damped power spectra at all radii \citep{benson_dark_2013,schneider_halo_2013}; however, subsequent work by \myref{leo_new_2018} demonstrates that this over-suppresses the production of small haloes. They find that using a smoothed version of the sharp \emph{k}-space filter produces halo mass functions in best agreement with \Nbody{} simulations. Throughout this paper, we use the \myref{leo_new_2018} smooth \emph{k}-space filter for the \WDM{} models that we consider.

To obtain our estimates of the number of substructures, \Nsub{}, within \RNFW{} of MW-like haloes we follow the approach described by \myref{giocoli_analytical_2008} that was subsequently modified in \extsec{4.4} of \myref{schneider_structure_2015} for use with sharp \emph{k}-space filters. Using the \myref{leo_new_2018} filter, a conditional halo mass function, $N_{\rm SK}$, is generated from the primordial linear matter power spectrum. \Myref{bode_halo_2001} showed that \WDM{} power spectra, $P_{\rm \WDM{}}\!\left(k\right),$ are related to the \CDM{} power spectrum, $P_{\rm \CDM{}}\!\left(k\right),$ by $P_{\rm \WDM{}}\!\left(k\right)=T^2\!\left(k\right) P_{\rm \CDM{}}\!\left(k\right),$ where $T\!\left(k\right)$ is the transfer function given by
\begin{equation}
    \label{eq:WDM_constraint_method:Methods:EPS:T_k}
    T\!\left(k\right) = \left[1 + \left(\alpha k\right)^{2\nu}\right]^\frac{-5}{\nu}.
\end{equation}
Here, $\nu=1.12$ and $\alpha$ is described by \myref{viel_constraining_2005} as being a function of the \WDM{} particle mass, \Mth{}, given by
\begin{equation}
    \label{eq:WDM_constraint_method:Methods:EPS:alpha}
    \alpha = 0.049\, \left[\frac{\Mth{}}{\keV{}}\right]^{-1.11}\,
                     \left[\frac{\upOmega_{\rm \WDM{}}}{0.25}\right]^{0.11}\,
                     \left[\frac{h}{0.7}\right]^{1.22} \cMpc{}.
\end{equation}
\Myref{schneider_structure_2015} showed that integrating the conditional halo mass function over the redshift-dependent spherical collapse threshold of a given progenitor, $\delta_c\!\left(z\right)$, gives the subhalo mass function
\begin{equation}
    \fdv{\Nsub{}}{\ln M} = \frac{1}{N_{\rm norm}} \int_{\delta_c\!\left(0\right)}^\infty
    \fdv{N_{\rm SK}}{\ln M}\, \dv{\delta_c}\,,
\end{equation}
where $M$ is the filter mass and $N_{\rm norm}$ is a normalization constant. The latter term, which is a free parameter, corrects the total count for progenitor subhaloes that exist at multiple redshifts which are counted more than once. Using the \myref{leo_new_2018} filter introduces two other free parameters, $\hat{\beta}$ and $\hat{c}$, that control the `smoothness' and the mass-radius relationship of the filter function.

\begin{figure}%
    \centering%
	\includegraphics[width=0.5\columnwidth]{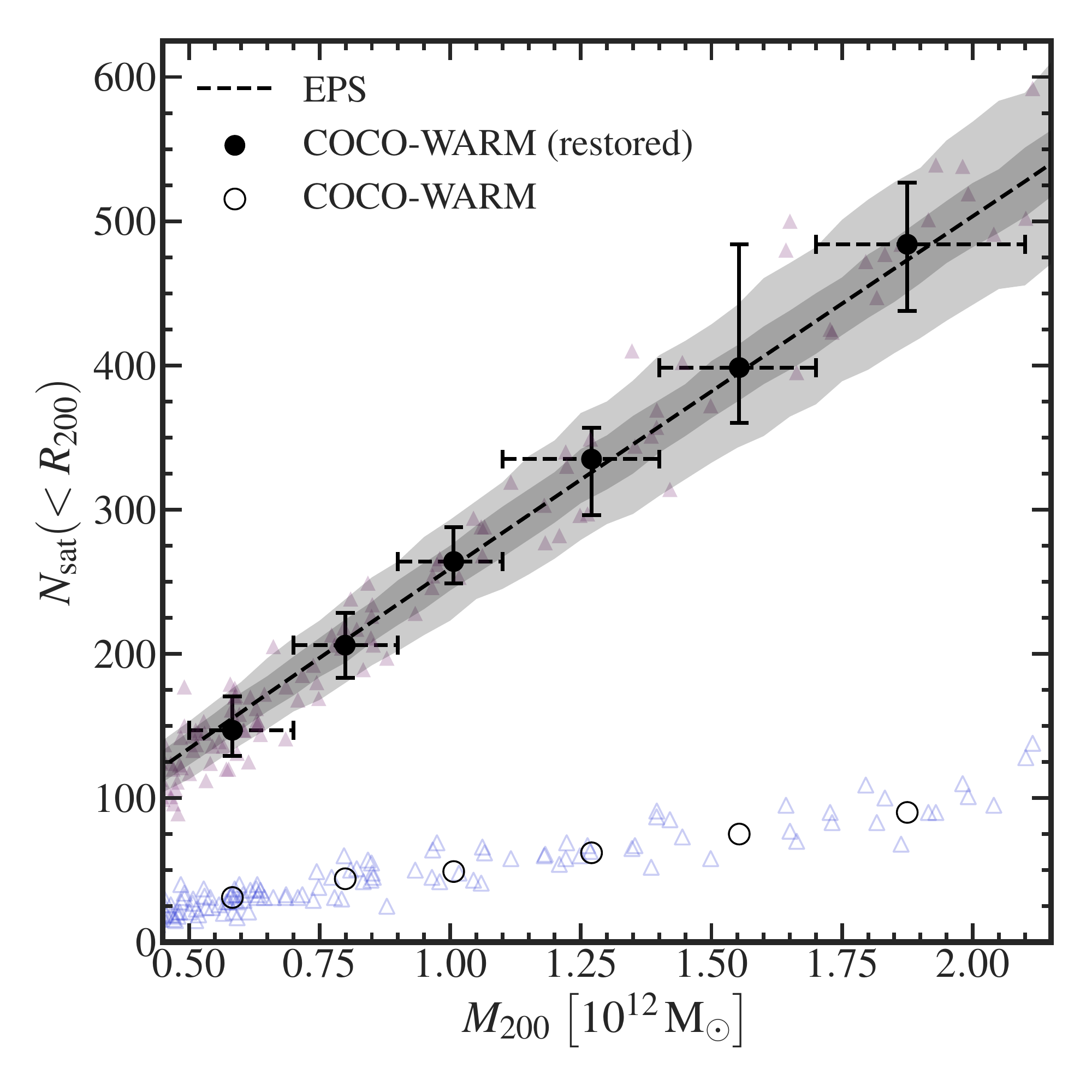}%
	\vspace{-10pt}%
	\caption{The total number of DM subhaloes within \RNFW{} as a function of DM halo mass, \MNFW{}. The dashed line shows the mean number of subhaloes predicted by the EPS formalism and the dark shaded region indicates the associated \percent{68} Poisson scatter. The light shaded region gives the \percent{68} scatter modelled using the negative binomial distribution given by \eqnref{eq:WDM_constraint_method:Methods:EPS:Neg_binom}. Triangular symbols represent individual haloes from the \wCOCO{} simulations and circular symbols represent the mean of the number of subhaloes in haloes in each mass bin. The width of each halo mass bin is indicated by a horizontal dashed error bar and the vertical error bar displays the corresponding \percent{68} scatter. In both cases, unfilled symbols represent objects from a subhalo catalogue where the `prematurely destroyed' subhaloes have not been recovered, and filled symbols indicate the same haloes using the subhalo catalogue after restoration of the `prematurely destroyed' subhaloes.
 	}%
	\label{fig:Methods:calibrate_EPS}%
	\vspace{-10pt}%
\end{figure}

We calibrate the free parameters of the EPS formalism by comparing its predictions of DM substructure with the fiducial subhalo populations of \COCO{} haloes in the mass bin\linebreak
${\MNFW{} = \left[0.95,\, 1.10\right]\Msun[\times10^{12}]}$. Specifically, we determine the EPS free parameters by applying the following two criteria:
\begin{enumerate}
    \item the EPS estimate of the mean number of \CDM{} subhaloes with mass $M \geq \Msun[10^9]$ must equal the mean number of objects with $\Mpeak{}\geq\Msun[10^9]$ in \cCOCO{} haloes; and,
    \item the EPS prediction of the mean number of \WDM{} subhaloes with ${M \geq \Msun[10^6]{}}$ must equal the mean number of objects with $\Mpeak{}\geq\Msun[10^6]$ in \wCOCO{} haloes (i.e. all subhaloes).
\end{enumerate}
Here, \Mpeak{} is determined using the \Subfind{} definition of halo mass \citep{springel_populating_2001,dolag_substructures_2009} and represents the highest mass achieved by the subhaloes at any time during their evolutionary histories. Typically, haloes achieve \Mpeak{} just before infall into a more massive halo. In the second calibration criterion,
we compare the mass functions at \Msun[10^6] as this is below the turnover in the \WDM{} power spectrum used in \wCOCO{}. We obtain excellent agreement between the mean EPS estimates and the \COCO{} simulation results by setting ${N_{\rm norm}=2.59},\, {\hat{\beta}=4.6},$ and ${\hat{c} = 3.9}$. This is shown in \figref{fig:Methods:calibrate_EPS}, which is discussed below.

The EPS formalism predicts only the mean number of subhaloes in DM haloes of a given mass, and not the host-to-host scatter in the subhalo count. As we will discuss later, including this scatter is very important to obtain unbiased results and thus needs to be accounted for. We do this using the results of cosmological \Nbody{} simulations that have shown that the scatter in the subhalo mass function is modelled well by a negative binomial distribution \citep{boylan-kolchin_theres_2010,cautun_subhalo_2014}. This takes the form
\begin{equation}
\label{eq:WDM_constraint_method:Methods:EPS:Neg_binom}
    {\rm P}\left(N \right|\left. r,\, p \right) =
    \frac{\upGamma\!\left(N+r\right)}
         {\upGamma\!\left(r\right)\upGamma\!\left(N+1\right)}\, p^r\! \left(1 - p\right)^N\,,
\end{equation}
where \emph{N} is the number of subhaloes and $\upGamma\!\left(x\right){=}\left(x-1\right)!$ is the Gamma function. The variable, ${p=\langle N \rangle\, /\, \upsigma^2,}$ where $\langle N\rangle$ and $\upsigma^2$ are, respectively, the mean and the dispersion of the distribution. This scatter in the subhalo count can be described best as the convolution of a Poisson distribution with a second distribution that describes the additional intrinsic variability of the subhalo count within haloes of fixed mass, such that ${\upsigma^2=\upsigma^2_{\rm Poisson}+\upsigma^2_{I}}$. The parameter \emph{r} then describes the relative contribution of each of these two terms: ${r=\upsigma_{\rm Poisson}^2\, /\, \upsigma_{\rm I}^2}$. We find that the scatter in the subhalo count of haloes in the \COCO{} suite is modelled well by $\upsigma_I{=}0.12\langle N\rangle$, as depicted in \figref{fig:Methods:calibrate_EPS}. We use this approach to characterize the scatter associated with the EPS predictions throughout the remainder of this paper.

In \figref{fig:Methods:calibrate_EPS}, we compare the EPS predictions for haloes in the mass range ${\left[0.5,\,2.0\right]\times10^{12}\Msun{}}$ to the number of subhaloes in individual \COCO{} haloes of the same mass. We obtain excellent agreement with the \Nbody{} results across the entire halo mass range of interest for this study. In particular, our approach reproduces very well both the mean number of subhaloes and its halo-to-halo scatter, which are represented by the grey shaded region and the vertical error bars, respectively.

\subsection{Calculating model acceptance probability}
\label{sec:Methods:Calculate_acceptance_probability}

We rule out sections of the viable thermal relic \WDM{} parameter space by calculating the fraction, \fviable{}, of \WDM{} systems that have at least as many subhaloes as the total number of MW satellites. We denote with $p^{\rm EPS}$ the probability density function of the number of DM subhaloes predicted by the EPS formalism. Then, the fraction of haloes with $\Nsat[\rm MW]$ or more subhaloes is given by
\begin{equation}
    \label{eq:WDM_constraint_method:Methods:EPS:F_acc_DMO}
    \fviable{}\!\left(\Nsub{} \geq \Nsat[\rm MW] \right) = 
        \int_{\Nsat[\rm MW]}^{\infty} \dv{\Nsub{}}\, p^{\rm EPS}\!\left(\Nsub{}\right)\,.
\end{equation}
However, as we discussed in \secref{sec:Methods:Model-independent_LF}, the inferred total number of MW satellite galaxies is affected by uncertainties. We can account for these by marginalizing over the distribution of MW satellite counts, $p^{\rm MW}\!\left(\Nsat[\rm MW]\right)$. Combining everything, we find that the fraction of WDM haloes with at least as many subhaloes as the MW satellite count is given by
\begin{equation}
    \label{eq:WDM_constraint_method:Methods:EPS:F_acc}
    \fviable{} = \int_0^\infty d\Nsat[\rm MW] \left[p^{\rm MW}\!\left(\Nsat[\rm MW]\right) 
        \;\;
        \int_{\Nsat[\rm MW]}^{\infty} \dv{\Nsub{}}\, p^{\rm EPS}\!\left(\Nsub{}\right)\right]\,.
\end{equation}
While not explicitly stated, both the number of MW satellites and the number of subhaloes (e.g. see \figref{fig:Methods:calibrate_EPS}) depend on the assumed MW halo mass \citep{wang_missing_2012}, which is still uncertain at the \percent{20} level (e.g. \citep{wang_mass_2020}). This means that the fraction of valid \WDM{} haloes depends strongly on the assumed mass of the Galactic halo. Note that the inferred total number of MW satellites depends weakly on the MW halo mass when calculated within a fixed physical distance, e.g. within 300\kpc{} from the Galactic Centre (see \extfig{10} in \myref{newton_total_2018}); however, here we calculate the expected number of satellites within \RNFW{} for each MW halo mass. 

\begin{figure}%
    \centering%
	\includegraphics[width=0.5\columnwidth]{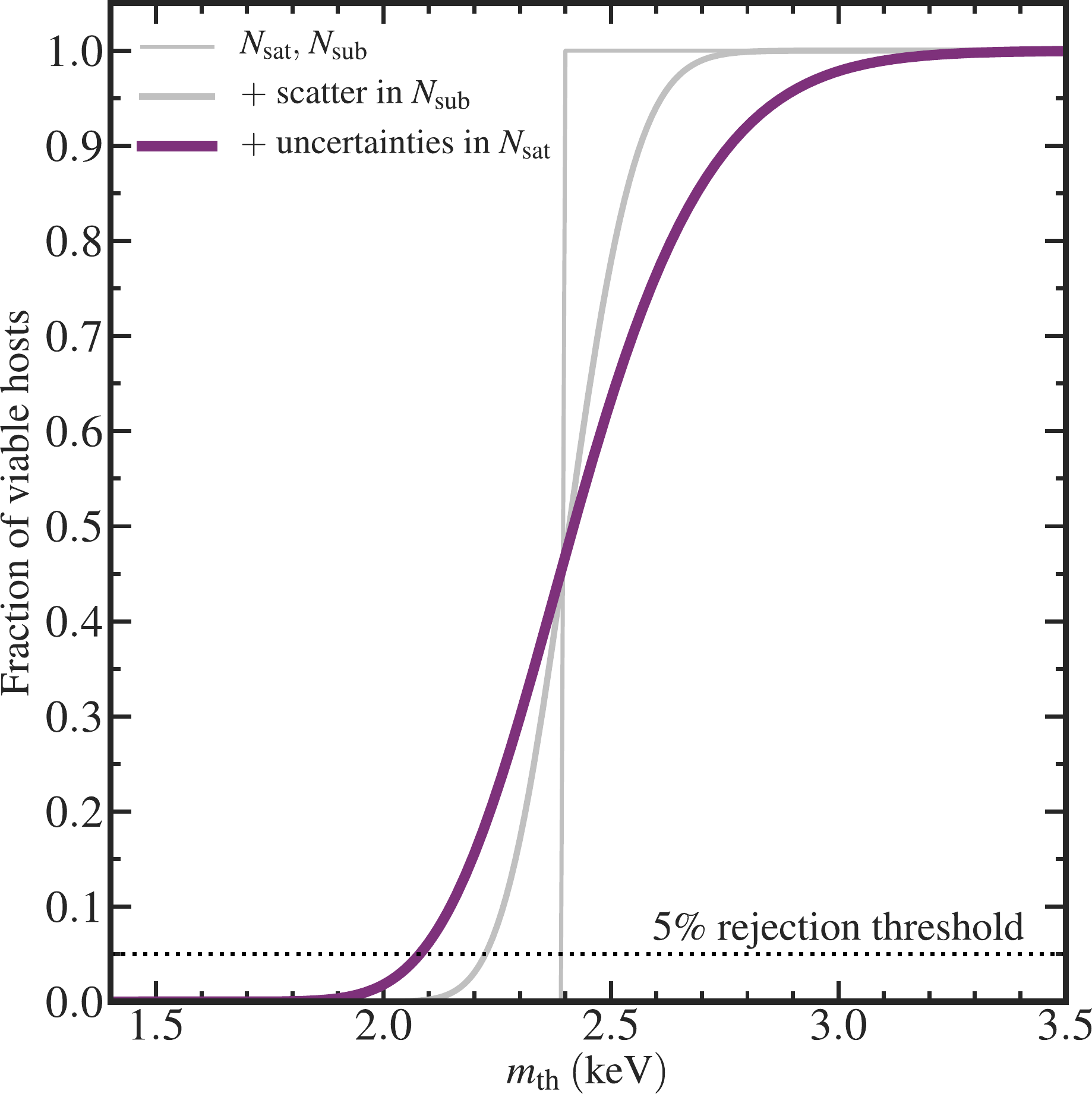}%
	\vspace{-10pt}%
	\caption{The fraction, \fviable{}, of \WDM{} systems with at least as many DM subhaloes, \Nsub{}, as the inferred total number of MW satellites, \Nsat{}, for a DM halo with ${\MNFW{}=\Msun[1\times10^{12}]}$. Thermal relic masses for which  $\fviable{}\leq0.05$ are ruled out with \percent{95} confidence. Earlier works that do not account for the uncertainty in \Nsat{} or the scatter in \Nsub{} at fixed halo mass (thin lines) artificially exclude too many thermal relic particle mass values. In this work (thick line) we include both sources of uncertainty in our calculation. The horizontal dotted line indicates the \percent{5} rejection threshold that we use to rule out parts of the \WDM{} parameter space.
	}%
	\label{fig:Methods:accepted_fractions}%
	\vspace{-10pt}%
\end{figure}%

This approach to calculating the fraction of viable \WDM{} systems for the first time incorporates the scatter in \Nsub{} at fixed halo mass \emph{and} the uncertainty in the inferred total MW satellite population. This is important, as excluding one, or both, of these sources of uncertainty produces constraints on \Mth{} that are too strict. We demonstrate this in \figref{fig:Methods:accepted_fractions} where, for each \WDM{} particle mass, we plot the fraction of haloes with mass $\MNFW[{\Msun[10^{12}]}]$ that contain enough DM substructure to host the inferred population of MW satellite galaxies. We derive our constraints on \Mth{} from the intersection of these cumulative distributions with the \percent{5} rejection threshold indicated by the horizontal dotted line. In this example, neglecting both sources of uncertainty excludes thermal relic DM with particle masses \MConstraint{2.4}, which is \percent{{\sim}15} more restrictive than our reported value of $\Mth{}{\lesssim}\keV[2.1]$ (thickest solid line). Some previous analyses (e.g.  \citep{polisensky_constraints_2011,lovell_properties_2014}) account for some of the uncertainty by modelling the scatter in the number of DM subhaloes at fixed halo mass. This weakens the constraint; however, the results are still artificially \percent{{\sim}5} more stringent than they should be with our more complete treatment of the uncertainties.
In addition to these complications, earlier works also suffer from incompleteness in the \z[0] subhalo catalogues due to numerical resolution effects. This contributes to a much more significant overestimation of the constraints and we discuss this in detail in the next section.

\section{Constraints on the thermal relic mass}
\label{sec:Constraints_on_mth}
Here we present the results of our analysis obtained using the EPS formalism calibrated to fiducial subhalo populations from the \cCOCO{} and \wCOCO{} simulations. Our most robust result assumes that \emph{all} DM subhaloes that form host a galaxy, thereby making no assumptions at all about galaxy formation processes. 

\subsection{Thermal relic particle mass constraints}
\label{sec:Constraints_on_mth:Constraint}

\begin{figure}%
    \centering%
	\includegraphics[width=0.5\columnwidth]{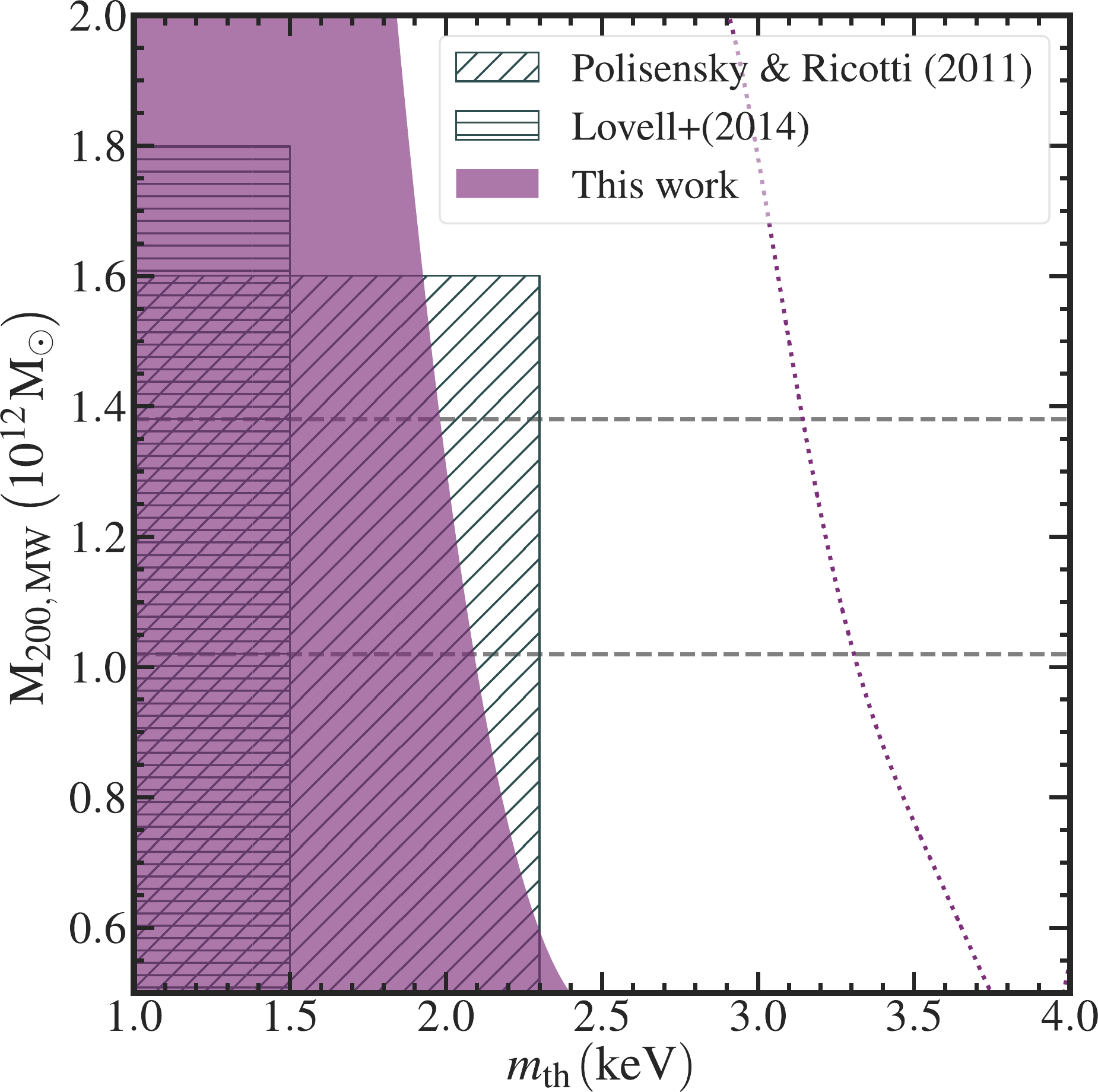}%
	\vspace{-10pt}%
	\caption{Constraints on the particle mass, \Mth{}, of the thermal relic \WDM{}. These depend on the assumed mass of the MW halo, which is shown on the vertical axis. We exclude with \percent{95} confidence parameter combinations in the shaded region. The dotted line indicates the extent of this exclusion region if we do not include `prematurely destroyed' subhaloes when calibrating the EPS formalism with the \COCO{} simulations~(see \secref{sec:Methods:EPS} for details). The constraints obtained by previous works, which do not consider some of the highest MW halo masses displayed here, are indicated by the hatched regions. These rule out too much of the parameter space as they do not account for some sources of uncertainty~(see \secref{sec:Methods:Calculate_acceptance_probability} for details). The two dashed horizontal lines show the \percent{68} confidence range on the mass of the MW halo from \myref{callingham_mass_2019}.
	}%
	\label{fig:Results:Thermal_exclusion_region}%
	\vspace{-10pt}%
\end{figure}%

We calculate the model acceptance distributions of DM haloes in the mass range\linebreak $\MNFW[{\left[0.5,2.0\right]\times\Msun[10^{12}]}]$ for several thermal relic \WDM{} models.  We rule out with \percent{95} confidence all combinations of \MNFW{} and \Mth{} with ${\fviable{} \leq 0.05}$. Problems arising from resolution effects persist even when using high-resolution simulations, and these effects are amplified as the resolution decreases. In addition to incorporating the scatter in \Nsub{} and the uncertainty in \Nsat{}, we account for resolution effects in the \Nbody{} simulations with which we calibrate the EPS formalism by including subhaloes that have been lost below the resolution limit at higher redshifts or destroyed artificially by tracking the most bound particle of these objects to \z[0] (for details see \appref{app:Convergence_tests}).

The results that we obtain using this approach are displayed in \figref{fig:Results:Thermal_exclusion_region}. The shaded region represents the parameter combinations that we rule out with \percent{95}~confidence. Independently of MW halo mass, we find that all thermal relic models with particle mass \MthTopConstraint{} are inconsistent with observations of the MW satellite population. The exact constraints vary with the MW halo mass, such that for lower halo masses we exclude heavier DM particle masses.

Recent studies, especially using \emph{Gaia} mission data, have provided more precise measurements of the MW halo mass (for a recent review, see \citep{wang_mass_2020}). We can take advantage of these results to marginalise over the uncertainties in the MW halo mass. For this, we use the \myref{callingham_mass_2019} estimate of the MW mass, which we illustrate in \figref{fig:Results:Thermal_exclusion_region} with two horizontal dashed lines indicating their \percent{68} confidence interval. This estimate is in good agreement with other MW mass measurements, such as estimates based on the rotation curve or on stellar halo dynamics \citep{cautun_milky_2020,wang_mass_2020}. Marginalising over the MW halo mass, we rule out all models with \marginalisedDMConstraint{}. These constraints do not depend on uncertain galaxy formation physics and therefore they are the most robust constraints to be placed on the thermal relic particle mass to date. A more realistic treatment of galaxy formation processes --- the effect of which would be to render a large number of low-mass subhaloes invisible --- would allow us to rule out more of this parameter space as fewer \WDM{} models would produce a sufficient number of satellites to be consistent with the inferred total population. We consider this possibility in more detail in \secref{sec:Galaxy_formation:Modelling_galform}.

In \figref{fig:Results:Thermal_exclusion_region}, we include for comparison the constraints obtained by \myrefs{lovell_properties_2014}{polisensky_constraints_2011} who use similar analysis techniques. These constraints suffer from resolution effects that suppress the identification of some substructures that survive to the present day. The dotted line demarcates the exclusion region that we would obtain in our analysis if we \emph{did not} account for these prematurely destroyed subhaloes. Such issues are not revealed by numerical convergence tests that are typically used to assess the reliability of particular simulations. For example, even the `level~$2$' simulations of \Aquarius{} haloes, which are some of the highest resolution DM-only haloes available, are not fully converged. We explore this in more detail in \appref{app:Convergence_tests}.

\section{The effects of galaxy formation processes}
\label{sec:Galaxy_formation}
\subsection{Modelling galaxy formation}
\label{sec:Galaxy_formation:Modelling_galform}
In the preceding sections we described an approach  that gives a highly robust, albeit conservative, lower limit on the allowed mass of the \WDM{} thermal relic particle. This ignores  the effects of galaxy formation processes on the satellite complement of the MW. These mechanisms play an important role in the evolution of the satellite galaxy luminosity function but still are not fully understood. Semi-analytic models of galaxy formation enable the fast and efficient exploration of the parameter space of such processes and thus help us to understand how they affect the \WDM{} constraints.

\Galform{} \citep{cole_recipe_1994,cole_hierarchical_2000} is one of the most advanced semi-analytic models that is currently available and is tuned to reproduce a selection of properties of the local galaxy population. A complete summary of the observational constraints used to calibrate the \Galform{} model parameters is provided in \extsec{4.2} of \myref{lacey_unified_2016}; hereafter \lacey{}. Of particular interest to our study is the reionization of the Universe, which is the main process that affects the evolution of the faint end of the galaxy luminosity function. The UV radiation that permeates the Universe (and that is responsible for reionization) heats the intergalactic medium and prevents it from cooling into low-mass haloes, impeding the replenishment of the cold gas reservoir from which stars would form.

In \Galform{}, the effect of reionization on haloes is modelled using two parameters: a circular velocity cooling threshold, \vcut{}, and the redshift of reionization, \zreion{}. The intergalactic medium is taken to be fully ionized at a redshift, \z[\zreion{}], whereafter the cooling of gas into haloes with circular velocities, ${v_{\rm vir} < \vcut{}}$, is prevented. This simple scheme has been verified against more sophisticated calculations of reionization, with which it has been shown to produce a good agreement \citep{benson_effects_2002,font_population_2011}. Recent studies by e.g.~\myref{bose_imprint_2018} have characterized the sensitivity of the satellite galaxy luminosity function to changes in these parameters: a later epoch of reionization allows more faint satellites to form, and a smaller circular velocity cooling threshold permits those faint satellites to become brighter.

We use \Galform{} to explore the effect of different parametrizations of reionization on the number of substructures containing a luminous component around the MW. Several previous works that have adopted a similar approach \citep{kennedy_constraining_2014,jethwa_upper_2017} used the \lacey{} model, which has \zreion[10] and \vcut[30]; however, this combination of parameters is now disfavoured by more recent theoretical calculations and the analysis of recent observational data, e.g. \citep{mason_universe_2018,planck_collaboration_planck_2020}. Additionally, others have noted that using \zreion[10] is not self-consistent and that a modified \lacey{} model with \zreion[6] is a more appropriate choice \citep{bose_imprint_2018}. In light of these theoretical and observational developments, for this study we consider parametrizations of reionization in the ranges $6 \leq \zreion{} \leq 8$ and $\kms[25] \leq \vcut{} \leq \kms[35]$ (see  \citep{okamoto_mass_2008,font_population_2011,robertson_cosmic_2015,banados_800-million-solar-mass_2018,davies_quantitative_2018,mason_universe_2018,planck_collaboration_planck_2020}).

\subsection{Constraints using GALFORM models}
\label{sec:Thermal_relic_constraints:Galform_constraints}

\begin{figure}%
    \centering%
    \includegraphics[width=0.5\columnwidth]{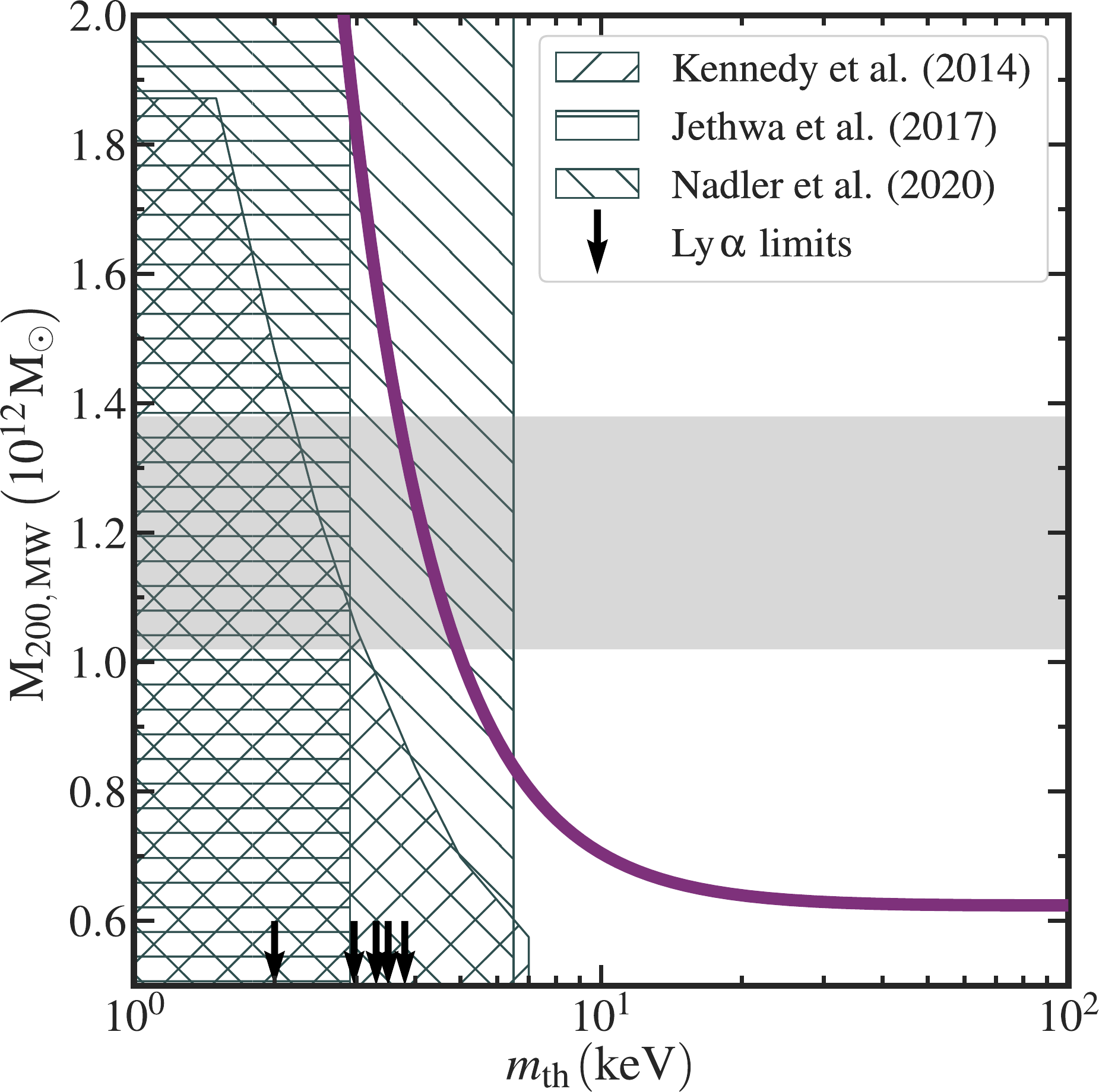}%
    \vspace{-10pt}%
    \caption{Constraints on \Mth{} obtained assuming our fiducial model of reionization with \zreion[7] and \vcut[30] within the \Galform{} galaxy formation model (thick solid line). Parameter combinations to the left of and beneath this envelope are ruled out with \percent{95} confidence. The constraints obtained by previous works that adopted similar approaches are displayed by the hatched regions \citep{kennedy_constraining_2014,jethwa_upper_2017,nadler_milky_2020}. Arrows indicate the \keV[2]~\citep{safarzadeh_limit_2018}, \keV[2.96]~\citep{baur_lyman-alpha_2016}, \keV[3.3]~\citep{viel_warm_2013}, \keV[3.5]~\citep{irsic_new_2017}, and \keV[3.8]~\citep{hsueh_sharp_2020} envelopes of the most robust constraints on the thermal relic particle mass obtained from modelling of the \Lyman{\upalpha} forest. The shaded region shows the \percent{68} confidence interval on the mass of the MW halo from \myref{callingham_mass_2019}.}%
    \label{fig:Results:Fiducial_Galform_constraint}%
    \vspace{-10pt}%
\end{figure}%

Our exploration of different prescriptions for reionization assumes the \lacey{} \Galform{} model as a reasonable description of various feedback and evolutionary processes in galaxy formation. We vary the reionization parameters in the ranges described in \secref{sec:Galaxy_formation:Modelling_galform} and apply \Galform{} to Monte Carlo merger trees calibrated as closely as possible to the \COCO{} suite.
The Monte Carlo algorithm used in \Galform{} cannot be calibrated to match exactly the \Nbody{} results as it lacks sufficient free parameters to match both the high- and low-mass ends of galaxy formation. Where a discrepancy exists between the Monte Carlo and \Nbody{} luminosity functions, we remap the \MV{} values of Monte Carlo satellite galaxies to new values such that the resulting luminosity function is consistent with the \Nbody{} results.
Using these, we obtain predictions for the dwarf galaxy luminosity function for $500$ realizations of each MW halo mass, allowing us to compute the model acceptance distributions in the same manner as before (see \secref{sec:Methods:Calculate_acceptance_probability}). Details of the merger tree algorithm and the functions to remap the Monte Carlo satellite galaxy $V-$band magnitudes are provided in \appref{app:PCH_algorithm_adjustment}.

\begin{figure}%
	\includegraphics[width=\textwidth]{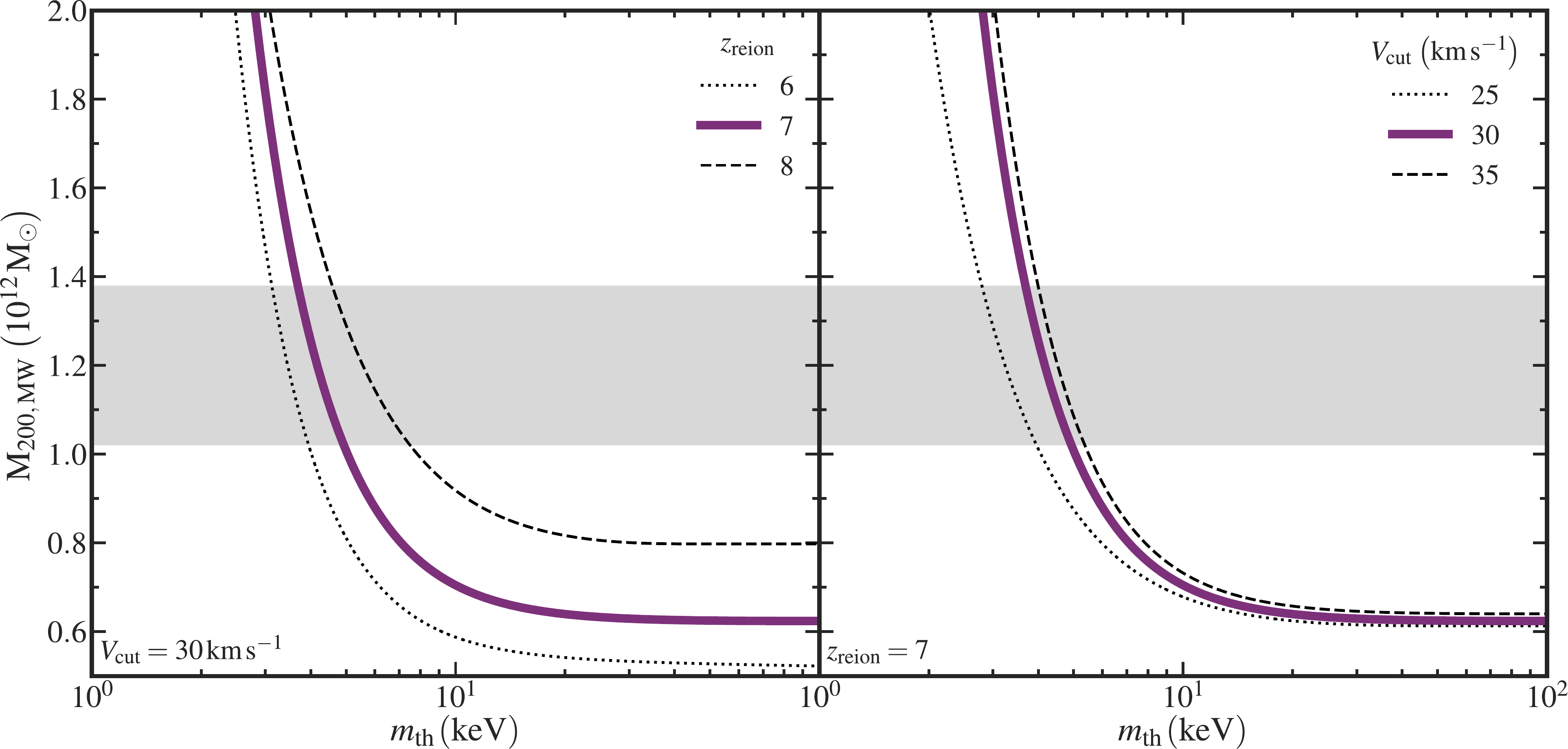}%
	\vspace{-10pt}%
	\caption{Constraints on \Mth{} obtained assuming different parametrizations of reionization in the \Galform{} galaxy formation model. Combinations of \MNFW{} and \Mth{} to the left of and beneath the envelopes are ruled out with \percent{95} confidence. In both panels, our fiducial choice is indicated by the thick solid line; the shaded region represents the \percent{68} confidence interval on the mass of the MW halo from \myref{callingham_mass_2019}. \emph{Left panel:} here, the cooling threshold is fixed at \vcut[30] and the dotted, solid, and dashed lines represent constraints obtained assuming $\zreion[6],\, 7,$ and $8$, respectively. High values of \zreion{} produce more stringent constraints on the thermal relic mass at fixed MW halo mass.
	\emph{Right panel:} here, reionization is assumed to have ceased by \zreion[7], and the dotted, solid, and dashed lines represent the constraints obtained assuming cooling thresholds of $\vcut[25,\, 30,\, {\rm and}\, 35],$ respectively. Higher cooling thresholds produce more stringent constraints on the thermal relic mass.}%
	\label{fig:Results:Galform_fixed_vcut_zcut}%
	\vspace{-10pt}%
\end{figure}%

In \figref{fig:Results:Fiducial_Galform_constraint}, we plot our constraints on thermal relic \WDM{} models assuming a fiducial model of reionization with \zreion[7] and \vcut[30]. This is a viable parametrization that is consistent with observations and resides in the centre of the parameter ranges that we explore. In this model, we rule out \emph{all} thermal relic \WDM{} particle masses with \MthGFTopConstraint{} independently of the MW halo mass. When marginalising over the uncertainties in the estimate of the MW halo mass from \myref{callingham_mass_2019}, our constraints strengthen and we exclude with 95 percent confidence all models with \marginalisedGalformConstraint{}. Our fiducial constraints are considerably stronger than our model-independent result and produce more stringent constraints in different MW halo mass regimes compared with work by refs. \citep{kennedy_constraining_2014,jethwa_upper_2017}, who also model the effects of galaxy formation processes.
More recently, \myref{nadler_milky_2020} carried out a similar analysis and obtained tighter constraints on the \WDM{} particle mass than we find in this work. We discuss the reasons behind this and its implications in \secref{sec:Discussion}.
In \figref{fig:Results:Fiducial_Galform_constraint}, we have also included for comparison the most conservative constraints derived from the \Lyman{\upalpha} forest by refs. \citep{viel_warm_2013,baur_lyman-alpha_2016,irsic_new_2017,safarzadeh_limit_2018,hsueh_sharp_2020}, which our results complement.

In \figref{fig:Results:Galform_fixed_vcut_zcut}, we explore the effect on the constraints of varying \vcut{} or \zreion{} while holding the other parameter constant. The left panel shows the effect of varying the redshift at which reionization concludes while fixing \vcut[30]. An epoch of reionization that finishes later, characterized by a lower value of \zreion{}, allows more faint galaxies to form in low-mass DM haloes, which weakens the constraints that can be placed on thermal relic \WDM{} models. The right panel shows the effect of curtailing further star formation in low-mass haloes after reionization finishes at \zreion[7]. As the \vcut{} cooling threshold increases, a larger fraction of the low-mass galaxy population is prevented from accreting new cold gas from the intergalactic medium after the end of reionization. Consequently, the reservoir of cold gas available for further star formation in these galaxies depletes over time, limiting how bright these objects become by \z[0]. When the cooling threshold is large, fewer faint galaxies evolve to become brighter than \MV[0] and populate the MW satellite galaxy luminosity function, leading to stronger constraints on the thermal relic mass. For completeness, in \appref{app:Galform_model_results} we provide the constraints obtained for the three values of \vcut{} assuming two scenarios with $\zreion[{6\, {\rm and}\, 8}]$, respectively.

\section{Discussion}
\label{sec:Discussion}
We have placed new conservative and highly robust constraints on the mass of the thermal relic \WDM{} particle by comparing EPS predictions of the DM subhalo content of \WDM{} haloes with the total number of MW satellite galaxies inferred from observations. We obtain estimates of the total satellite complement using the \myref{newton_total_2018} approach including recent observations of satellites from the \SDSS{} and \DES{}. To calibrate the EPS formalism, we use DM haloes from the \COCO{} simulation suite with masses in the likely MW halo mass range $\MNFW{}{=}\Msun[{\left[0.5,\, 2.0\right]\times10^{12}}]$. We improve upon previous constraints by incorporating for the first time the uncertainty in the size of the total MW satellite population and by accounting for unresolved or numerically disrupted subhaloes in \Nbody{} simulations (see \appref{app:Convergence_tests}). In a separate analysis we also explore the effect of various assumptions about galaxy formation processes on the constraints that we can place on the \WDM{} particle mass.

We find that, when marginalizing over uncertainties in estimates of the MW halo mass, thermal relic models with \marginalisedDMConstraint{} are ruled out with \percent{95} confidence (see \figref{fig:Results:Thermal_exclusion_region}). This result is independent of assumptions about galaxy formation physics, as for our purposes we treat \emph{all} DM subhaloes as hosts of visible galaxies. This ensures that the constraints provide a robust lower limit on the mass of the thermal relic \WDM{} particle, improving on the results reported in \myref{lovell_properties_2014} across the entire MW halo mass range considered (see \figref{fig:Results:Thermal_exclusion_region}). Our results are competitive with but slightly less restrictive than the constraints obtained by \myref{polisensky_constraints_2011} because we account for subhaloes that exist but are missing for numerical reasons from the \z[0] halo catalogues.

The resolution of a simulation can affect the population of haloes at \z[0] in two major ways. First, haloes close to the resolution limit of a simulation experience stronger tidal disruption due to numerical effects that can destroy the halo. Secondly, some structure finders stop tracking haloes that fall below a mass threshold at \emph{any} time during their evolution. Haloes composed of few particles can occasionally fall below this and recover later, with the result that the object is permanently excluded from the final catalogue even if it survives to the present day. Omitting these objects significantly affects the constraints on the \WDM{} parameter space, strengthening them artificially (see \figref{fig:Results:Thermal_exclusion_region}). This effect worsens as simulation resolution decreases, so constraints that are obtained using lower-resolution simulations and methods that do not account for `prematurely destroyed' subhaloes will be significant overestimates.

The processes responsible for the formation of galaxies are complex and are yet to be understood fully; nevertheless, they play an important role in shaping the luminosity function of the dwarf galaxies of the MW. Incorporating the effects of these mechanisms into our approach allows us to refine the constraints on the properties of the DM and rule out many more \WDM{} models. In a modified version of the \lacey{} \Galform{} model with \zreion[7] and \vcut[30] (our fiducial model) we rule out, with \percent{95} confidence, thermal relic models with \marginalisedGalformConstraint{} when marginalizing over uncertainties in the MW halo mass (see \figref{fig:Results:Fiducial_Galform_constraint}). Furthermore, we rule out all thermal relic \WDM{} particle masses with \MthGFTopConstraint{} independently of MW halo mass. These improve on our model-independent results and are consistent with the constraints obtained in previous works that adopted similar approaches. This result also compares favourably with complementary constraints derived from the \Lyman{\upalpha} forest by refs.~\citep{viel_warm_2013,baur_lyman-alpha_2016,irsic_new_2017,safarzadeh_limit_2018,hsueh_sharp_2020}.

Recently, \myref{nadler_milky_2020} conducted a similar analysis to constrain the particle mass of thermal relic \WDM{} using the inferred luminosity function of MW satellite galaxies from
\myref{drlica-wagner_milky_2020}.
Their constraints on the DM particle mass are stricter than all of our results spanning the parametrizations of reionization considered in this work (see \figrefs{fig:Results:Galform_fixed_vcut_zcut}{fig:Appendix:Galform_constraints:non_fid_z}, and \tabref{tab:Appendix:Galform_constraints:galform_model_constraints}).
Two factors contribute to this discrepancy. First,
\myref{nadler_milky_2020} use an abundance matching technique extrapolated to very faint magnitudes to populate substructure with galaxies. Such techniques adopt a model to describe the relationship between the DM structure and the luminous component; however, they may not capture the full complexity of galaxy formation physics at the faint end \citep{sawala_bent_2015}. Semi-analytic models like the one used in this work are physically motivated and fare better at modelling the baryonic processes taking place on small scales, encapsulating more of the complexities of galaxy formation in this regime; however, they are not entirely free of simplifying assumptions. Secondly, the \myref{nadler_milky_2020} results are based on a combination of \PAN{} and \DES{} data whereas our analysis uses satellite galaxy data from \SDSS{} and \DES{}. The \PAN{} survey data are not as deep as those from \SDSS{}, particularly at the faint end of the satellite galaxy luminosity function. Consequently, the size of the satellite population inferred from the \PAN{} data, and hence the \WDM{} constraint derived from this, is more sensitive to modelling uncertainties in the inner halo.
The discrepancy in the DM particle mass constraints between these two approaches demonstrates the role that uncertainties in galaxy formation physics play in analyses of this type and motivates continued efforts to further our understanding of these processes. It also shows that the incompleteness of existing surveys of the MW satellite galaxy population contributes to analysis uncertainties.
Future surveys such as the Legacy Survey of Space and Time~(\LSST{}) to be carried out by the Vera~C.~Rubin Observatory will improve the sky coverage and depth of extant surveys of the MW halo and help to tighten the uncertainties on DM particle mass constraints.

Two important aspects of the reionization of the Universe affect the formation of the low-mass galaxy population. The timing of the end of reionization influences how many low-mass DM haloes are able to accrete cold gas for use in star formation prior to reionization. The later reionization finishes, the more time is afforded for faint galaxies to form in such haloes. After this, further cold gas accretion is limited to those haloes that are massive enough that the gas can condense out of the intergalactic medium and onto the galaxy. The star formation that this facilitates enables the faintest galaxies to become brighter, changing the shape of the faint end of the satellite galaxy luminosity function \citep{bose_imprint_2018}. These processes are reflected in our constraints (see \figrefs{fig:Results:Galform_fixed_vcut_zcut}{fig:Appendix:Galform_constraints:non_fid_z}), where we find that an epoch of reionization that finishes earlier (i.e. at higher values of \zreion{}) and a larger cooling threshold (\vcut{}) produce the most stringent constraints on the thermal relic particle mass. At high MW halo mass well away from the lower limit of the constraint envelope, the value chosen for \vcut{} has the largest effect on the number of substructures with a luminous component, in agreement with previous work, e.g. refs.~\citep{kennedy_constraining_2014,jethwa_upper_2017}. However, close to the MW halo mass favoured by \myref{callingham_mass_2019}, we find that the choice of \zreion{} has a significant effect on the constraints that can be placed on thermal relic models.

Our key results (see \figref{fig:Results:Thermal_exclusion_region}) assume MW halo masses in the most likely range\linebreak
${\MNFW[{\left[0.5,\, 2.0\right]\times10^{12}\Msun{}}]}$. The constraints have only a moderate dependence on host halo mass because the number of MW satellite galaxies within a fixed radius inferred from observations scales much less strongly with halo mass than the number of subhaloes predicted by DM models (see \secref{sec:Methods:EPS}). Better measurements of the mass of the MW halo will improve the constraining power of this approach; in the most extreme case, a MW halo with mass at the lowest end of the likely range would rule out thermal relic models with \MConstraint{2.4} independently of galaxy formation physics. This estimate does not account for the effect of the central baryonic disc of the host halo that destroys subhaloes \citep{gnedin_tidal_1999,brooks_why_2014,garrison-kimmel_not_2017,sawala_shaken_2017,richings_subhalo_2020,richings_high-resolution_2021,webb_high-resolution_2020}, which would exclude more of the \WDM{} parameter space. For our fiducial galaxy formation model (see \figref{fig:Results:Fiducial_Galform_constraint}), we also find that all DM particle masses are excluded for MW halo masses, $\MNFW{}\leq\Msun[0.6\times 10^{12}]$. This arises from the failure of the models to produce enough faint galaxies to be consistent with observations of the MW satellites, even in very cold thermal relic models where the number of low-mass subhaloes does not differ significantly from \CDM{} predictions. Therefore, this threshold can be interpreted as a lower-mass limit for our Galaxy within the \CDM{} model (see also \citep{busha_mass_2011,cautun_milky_2014}).

Recently, the \Edges{} collaboration announced the detection of a global \cm[21] absorption line in measurements of the cosmic microwave background radiation \citep{bowman_absorption_2018}. This shows promise as a potential complementary probe of \WDM{} models at high redshift because its shape and location (\z[17.2]) depend partly on the abundance of low-mass structures that act as sites of early star formation \citep{dayal_reionization_2017}. Currently, this epoch is inaccessible to other observational techniques \citep{chatterjee_ruling_2019}. Unfortunately, the \cm[21] signal is very sensitive to uncertainties in the modelling of the Galactic foreground and in our understanding of the physics of star formation at early times. Therefore, the current data cannot constrain the properties of the DM \citep{boyarsky_21-cm_2019,leo_constraining_2020,rudakovskyi_can_2020}. Future studies of the statistics of the spatial distribution of the \cm[21] signal and further work to understand stellar evolution at high redshift will overcome these difficulties \citep{leo_constraining_2020,rudakovskyi_can_2020}.

The size of the satellite population inferred by the \myref{newton_total_2018} method is a lower limit to the true population as it cannot account for spatially-extended dwarf galaxies that fall below the surface brightness threshold of the surveys. Additionally, it does not encompass the contribution of the former satellites of the Large Magellanic Cloud that lie outside the \DES{} footprint that could increase the size of the satellite complement still further. Taken together, these caveats \emph{strengthen} the robustness of our lower limits on the thermal relic particle mass as a larger inferred satellite complement would rule out an even larger region of \WDM{} parameter space.

\section{Conclusions}
\label{sec:Conclusions}
In the continued absence of the direct detection of a DM particle or the observation of an astrophysical phenomenon that unambiguously constrains its properties, the debate about its exact nature and the acceptability of the current cosmological paradigm will continue. The discussion of `small-scale' challenges to \LCDM{} --- perceived discrepancies between the observations of low-mass galaxies and predictions of DM substructure --- has renewed impetus in this regard and has encouraged further exploration of alternative DM models. One class of these, which are broadly termed \WDM{} models, produces a cut-off in the linear matter power spectrum that leads to a suppression in the formation of DM haloes on the scale of (and smaller than) those that would usually host dwarf galaxies in \LCDM{}. The location and nature of this suppression depends sensitively on the properties of the DM particle. One method to constrain the parameter space of these models is the use of sophisticated hydrodynamic simulations to simulate self-consistently the formation and evolution of dwarf galaxies in the Local Group, and around MW-like hosts in particular. However, the resolution that would be required to achieve this in a volume that is large enough to attain high statistical power is, at present, computationally challenging. The development of other approaches to explore efficiently the viability of different cosmological models on these scales is, therefore, important. 

In this work, we improve a method to constrain the properties of \WDM{} models by comparing Extended Press-Schechter (EPS) predictions of the amount of substructure within MW-mass \WDM{} haloes with the most recent estimates of the size of the satellite population of the MW (see \secrefs{sec:Methods:EPS}{sec:Methods:Calculate_acceptance_probability}). This approach is complementary to previous work and for the first time accounts fully for limitations in the resolution of \Nbody{} cosmological simulations, incorporates the scatter in the number of substructures inside haloes at fixed DM halo mass, and includes the uncertainty associated with estimates of the number of satellite galaxies in the MW. The constraints that can be produced by this method are efficient at ruling out \WDM{} models independently of any particular choice of galaxy formation physics, making the results highly robust.

We demonstrate the utility of this approach by applying it to thermal relic \WDM{} models to constrain the DM particle mass (see \secref{sec:Constraints_on_mth:Constraint}). Our most robust constraint rules out, with \percent{95} confidence, thermal relic \WDM{} particles with masses \marginalisedDMConstraint{} when marginalizing over uncertainties in estimates of the MW halo mass. This is competitive with existing limits that also use the abundance of MW satellite galaxies to constrain the \WDM{} parameter space with minimal assumptions; however, our approach accounts for small subhaloes in \Nbody{} simulations that are not identified by substructure finders for numerical reasons, even though some of them actually survive to \z[0]. Excluding them from the subhalo catalogue reduces the number of subhaloes that are available to host dwarf galaxies, artificially strengthening restrictions on the viable thermal relic model parameter space (see \figref{fig:Results:Thermal_exclusion_region}). This effect worsens as the simulation resolution becomes poorer, so constraints that are obtained using lower-resolution simulations without accounting for the `prematurely destroyed' subhaloes are significant overestimates.

All methods that seek to constrain the properties of DM models using visible tracers of the underlying substructure must make assumptions about galaxy formation processes that affect the satellite complement of the MW. Here, to obtain our highly robust constraints on the allowed properties of candidate \WDM{} particles independently of galaxy formation physics, we have made the minimal and conservative assumption that a galaxy forms in \emph{all} DM haloes. This allows us to place stringent lower bounds on the parameter space of thermal relic \WDM{} models. In reality, baryonic physics mechanisms are important to determine the fraction of DM haloes that go on to host visible galaxies at late times, leaving many small subhaloes `dark' \citep{benitez-llambay_detailed_2020}. While the details of these processes are still not understood fully, they are now constrained quite well. Accounting for these physical processes in models reduces the effective size of the satellite complement and in our analysis this improves significantly the constraints on the \WDM{} particle properties.

Of particular interest to this study, the reionization of hydrogen in the early Universe, and the size of DM haloes in which it suppresses galaxy formation, dominates the formation and evolution of low-mass galaxies and imprints a characteristic signature on the luminosity function of MW satellite galaxies. We use the Durham semi-analytic model \Galform{} to explore several possible descriptions of this process and examine how different parametrizations affect the constraints on thermal relic \WDM{} (see \secref{sec:Galaxy_formation:Modelling_galform}). By assuming that reionization is complete by \zreion[7] and that galaxy formation is suppressed in DM haloes with circular velocity $v_{\rm vir}{<}\kms[30]$, we rule out with \percent{95} confidence thermal relic DM with mass \marginalisedGalformConstraint{}, when marginalizing over uncertainties in estimates of the MW halo mass (see \figref{fig:Results:Fiducial_Galform_constraint}). We also find that a MW halo mass below \MNFW[{\Msun[0.6\times 10^{12}]}] would not permit \emph{any} thermal relic models that are warmer than \CDM{}. This improves on the \myref{kennedy_constraining_2014} result and is competitive with conservative astrophysical limits from recent analyses using the \Lyman{\upalpha} forest. Furthermore, we find that the redshift at which reionization is assumed to cease has a significant effect on the constraints near to the most likely MW halo mass; however, for large MW halo masses the value chosen for the cooling threshold is more important (see \figref{fig:Results:Galform_fixed_vcut_zcut}). Continued efforts to constrain further the probable ranges of the reionization parameters and the mass of the MW are therefore crucial if we wish to place ever more stringent constraints on the viability of alternative models to the \LCDM{} paradigm.

While a DM particle candidate remains undetected, \WDM{} models remain a feasible alternative to \CDM{}. The satellite galaxy system of the MW provides a powerful means of probing structure formation on small scales and can help to discriminate between different cosmological models. However, the MW may not be typical of most DM haloes of similar mass. Hydrodynamic simulations that self-consistently model star formation and gas physics on the scale of dwarf galaxies will facilitate more robust astrophysical tests of this; however, achieving sufficient resolution is computationally challenging at present. A complementary means of testing the predictions of structure formation from different cosmological models is to consider their predictions of the evolution of structure across a range of mass scales and in a variety of environments, and to compare these with observations. Currently, this is challenging as it is difficult to identify the faintest and most extended objects at vast distances against observational backgrounds. Future improvements in observational capability will offer the prospect of further constraining the parameter space of viable \WDM{} models.



\appendix
\section{Resolution effects in numerical convergence studies}
\label{app:Convergence_tests}

Numerical simulations are a useful tool to study the physical behaviour of cosmological models in the non-linear regime, where analytical approaches are unable to account fully for the complexity at these scales. While the dynamic range of such simulations is vast, spanning many orders of magnitude, \Nbody{} simulations are limited by the resolution at which their smallest objects can be self-consistently modelled. It is important to understand whether the phenomena that are observed in the simulations occur for physical reasons, or whether they arise \emph{because} of this limitation.

The traditional approach to identify the onset of resolution effects has been to conduct convergence studies, e.g.~\citep{efstathiou_numerical_1985,power_inner_2003}. These entail re-running the same simulation at different resolution levels and comparing the results: those that are unaffected by an increase in the resolution are deemed to be converged. A number of studies using several different \Nbody{} simulations support this conclusion and suggest that the subhalo present-day mass function of DM haloes is converged down to approximately $100$ simulation particles per object, e.g.~\citep{springel_aquarius_2008,onions_subhaloes_2012,griffen_caterpillar_2016}. Some of these low-mass subhaloes could be disrupted by numerical effects from the limited resolution of the simulation \citep{van_den_bosch_dark_2018,green_tidal_2019,errani_asymptotic_2021,green_tidal_2021}. \Myref{onions_subhaloes_2012} also show that configuration space structure finders are ineffective at identifying all substructure near the centre of simulated haloes. This  resolution-dependent deficiency of the halo finding algorithms implies that some substructures may be missed. These effects complicate attempts to understand and characterize the convergence of the subhalo \emph{peak} mass function, which is of interest for this study as peak mass correlates more strongly with the formation of a luminous component than the present-day halo mass. It also affects directly the calibration of the EPS formalism that we use to estimate the amount of substructure in MW-mass haloes. 

In the peak mass function, resolution limitations can also affect the high-mass end as even haloes with large peak mass can be excluded from the \z[0] halo catalogue if they fall below the resolution limit. This could occur after many orbits of the host during which the subhalo experiences continuous tidal stripping of mass. It is important to correct for missing and `prematurely disrupted' subhaloes as these can bias our results: as we discuss in the main text, under-predicting the true subhalo count produces overly stringent constraints on the \WDM{} particle mass. We are also careful to distinguish these from the spurious haloes found in \Nbody{} \WDM{} simulations, which are produced by artificial fragmentation of filaments due to numerical effects and should be removed from the halo catalogues.

The `prematurely destroyed' subhaloes may be recovered relatively easily by tracing their constituent particles through the simulations and identifying whether they survive to the present day. Details may be found in \extapp{B} of \myref{newton_total_2018}.
Briefly, we use the \myref{simha_modelling_2017} merging scheme implemented in \Galform{} to
carry out this procedure. This tracks the most bound particle of objects that fall below the resolution limit from the last epoch at which they were associated with a resolved subhalo. From this, a population of substructures is recovered that contains the `prematurely destroyed' subhaloes and other objects that are disrupted by physical processes. We remove the latter from the recovered population if they satisfy one of the following criteria:
\begin{enumerate}
    \item A time has elapsed after the subhalo fell below the resolution limit, which is equal to or greater than the dynamical friction timescale.
    \item At any time, the subhalo passes within the halo tidal disruption radius.
\end{enumerate}
In both cases, the effects of tidal stripping and of interactions between orbiting subhaloes are ignored. The size of this correction to the \COCO{} suite is not easy to ascertain as \COCO{} does not have counterpart simulations with different resolution levels with which to conduct a similar convergence study. Instead, we use the \Aquarius{} suite \citep{springel_aquarius_2008}, the constituent simulations of which span a range of resolution levels that encompass that of \COCO{}, to estimate the size of the effect of excluding the prematurely destroyed subhaloes.

\begin{figure}%
    \centering%
	\includegraphics[width=0.5\columnwidth]{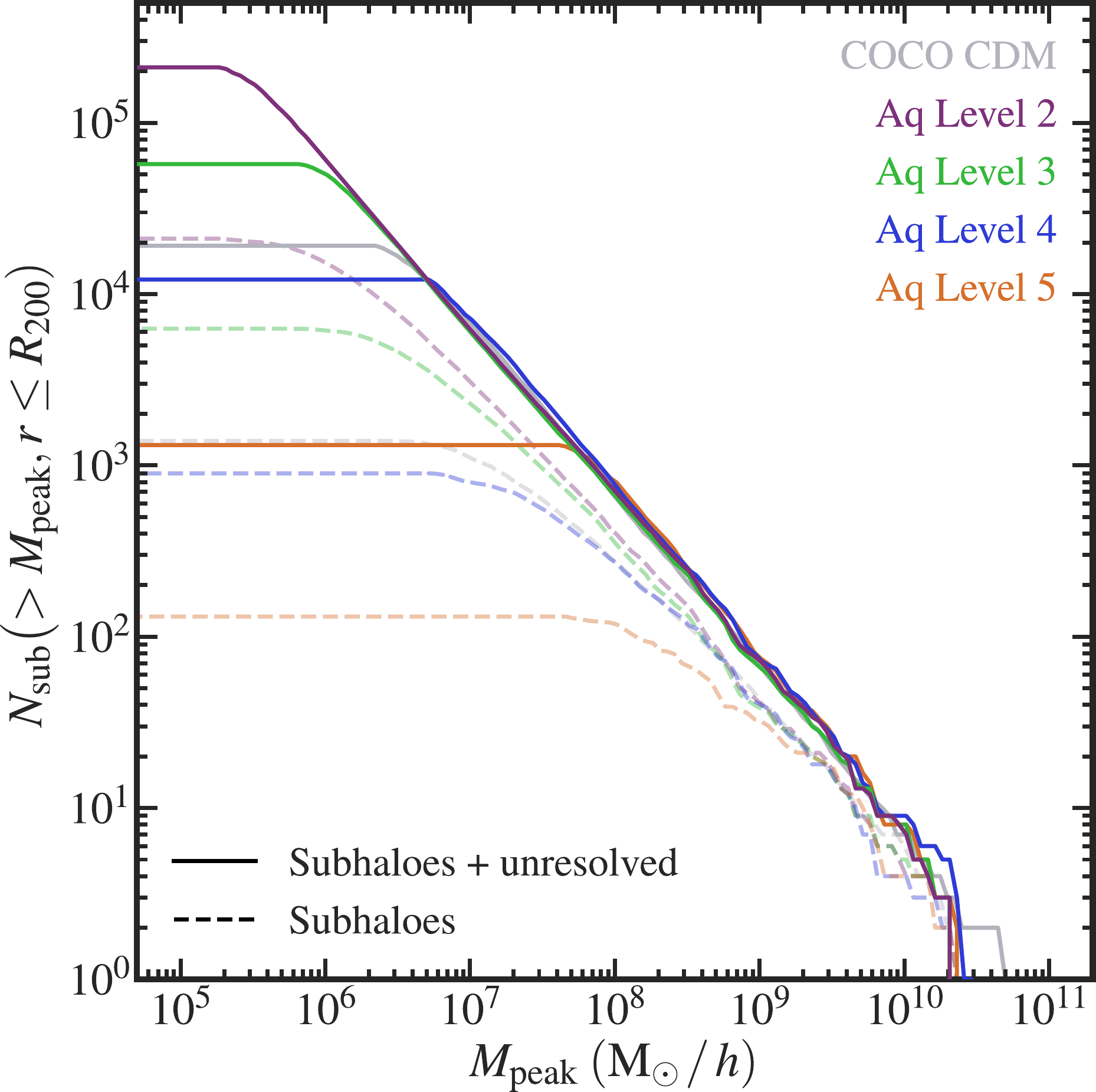}%
	\vspace{-10pt}%
	\caption{Cumulative subhalo peak mass functions of the \Aquarius{} A halo simulated at different levels of resolution (coloured lines) and stacked \cCOCO{} haloes (grey lines) with masses $\MNFW{}\geq\Msun[1.5\times10^{12}]$.
	The dashed lines show the original, uncorrected number counts prior to recovering the `prematurely destroyed' subhalo population. The solid lines show the number counts after adding this population to the original one.	The resolution level of the \COCO{} suite lies between \Aquarius{} Level~3 and Level~4.
	}%
	\label{fig:Appendix:Aq_convergence_test}%
	\vspace{-20pt}%
\end{figure}%

In \figref{fig:Appendix:Aq_convergence_test}, we compare the subhalo peak mass functions of the \Aquarius{} A halo simulated at four different resolution levels: 2, 3, 4, and 5. Aq~Level~5 is simulated coarsely, with a DM particle mass, $m_{\rm p}=\Msun[3.14\times10^6]$. The simulation resolution improves with decreasing level number, such that Aq~Level~2 is simulated with a DM particle mass, $m_{\rm p}=\Msun[1.37\times10^4]$ (i.e. a factor of ${\sim}200$ times better mass resolution). The figure shows the subhalo count before and after recovering the population of missing and prematurely destroyed subhaloes. At high halo mass, the original and `corrected' curves are consistent with the highest resolution simulation. As the resolution degrades, the lower-resolution simulations peel away from the Level~2 curves, with the lowest-resolution simulation turning off at the highest value of \Mpeak{}. This demonstrates the major consequence of limited resolution, which is particularly acute for low-mass objects: in the cases considered here for haloes with mass $\MNFW\geq\Msun[{1.5\times10^{12}}]$, restoring the missing population increases the total subhalo abundance by an order of magnitude. However, as we discussed earlier, resolution effects are not confined to the low-mass regime and can also affect higher masses. Massive haloes can experience considerable tidal stripping after being accreted by a host, which can lead to their exclusion from the \z[0] halo catalogue. The resulting discrepancy between the original and corrected mass functions at high masses indicates that this population of `missing' objects composes a non-negligible fraction of the subhaloes even in the high-mass regime. Therefore, `traditional' convergence studies that do not account for missing and prematurely destroyed subhaloes cannot properly characterize these numerical effects on the peak mass function.

In \figref{fig:Appendix:Aq_convergence_test}, we also plot for comparison the average mass function of \cCOCO{} haloes with masses similar to the \Aquarius{}~A halo. The \cCOCO{} and \wCOCO{} simulations have a DM particle mass resolution that lies between that of the Aq~Level~$3$ and Level~$4$ runs. Comparing the subhalo mass functions of the incomplete subhalo catalogues of \cCOCO{} and Aq~Level~3 suggests that subhaloes with $\Mpeak{} \gtrsim \Msun[3\times10^8]$ are resolved well. However, after recovering the prematurely destroyed subhaloes, a comparison of the mass functions implies consistency at masses $\Mpeak{} \gtrsim \Msun[5\times10^6]$, approximately two orders of magnitude better than before. This is consistent with the correction to the Aq~Level~4 simulation, which suggests that the same correction for prematurely disrupted subhaloes that we have shown to work well for the \Aquarius{} Level~2 to 5 runs is also applicable to the two \COCO{} simulations.
\section{Calibrating the Galform merger tree algorithm}
\label{app:PCH_algorithm_adjustment}

Monte Carlo merger trees are generated within \Galform{} using an implementation of the \myref{parkinson_generating_2008} merger tree algorithm, which iteratively splits the present-day halo mass into different progenitor haloes as it progresses to higher redshifts. The algorithm depends on three free parameters: $G_0{=}0.57$, a normalization constant; $\gamma_1{=}0.38$, which controls the mass distribution of the progenitor haloes; and $\gamma_2{=}-0.01$, which controls the halo-splitting rate. \Myref{parkinson_generating_2008} calibrated these parameters by comparing the Monte Carlo progenitor halo mass functions at several redshifts with those from the Millennium simulation \citep{springel_simulations_2005}. This follows the evolution of $2160^3$ particles with mass, $m_p = \cMsun[8.6\times10^8]$, resolving the halo mass function to $\cMsun[{\sim}1.7\times10^{10}]$, which is three orders of magnitude larger than the regime of interest for this study. The merger trees produced from the best-fitting free parameter values derived from the Millennium simulation predict a factor of two times more galaxies at the faint end of the cumulative luminosity function in MW-mass haloes than is obtained by applying \Galform{} to the \COCO{} suite.

\begin{figure}%
    \centering%
	\includegraphics[width=\textwidth]{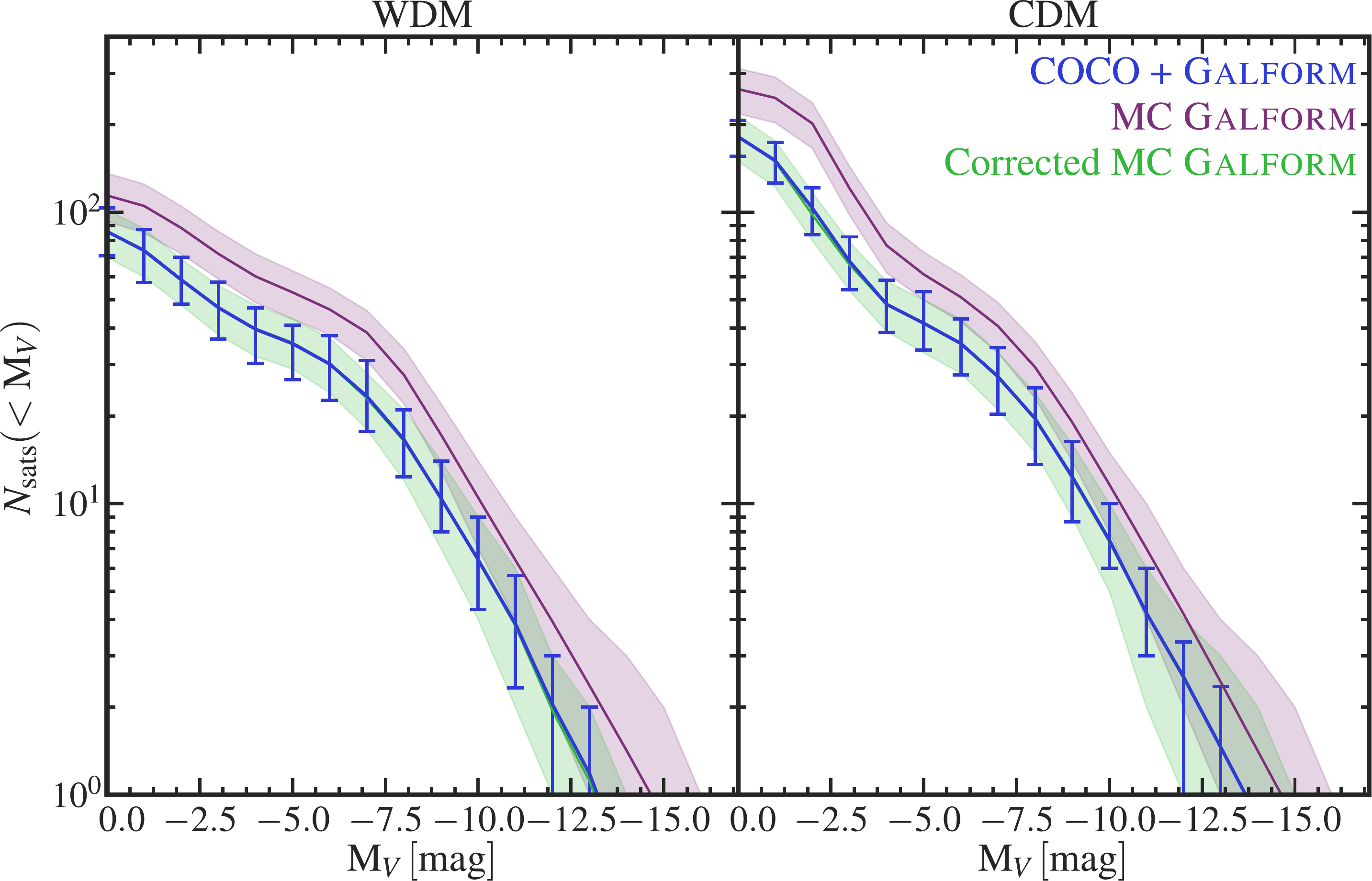}%
	\vspace{-10pt}%
	\caption{Cumulative satellite galaxy luminosity functions produced by our fiducial \Galform{} model with \zreion[7] and \vcut[30] for haloes with masses in the range ${\MNFW{}=\Msun[{\left[1,\, 1.5\right]\times10^{12}}]}$. Results for the \keV[3.3] thermal relic \WDM{} model and \CDM{} model are shown in the left and right panels, respectively.
	The mean luminosity functions produced from \Galform{} applied to the \COCO{} simulations are represented by blue solid lines and error bars, which indicate their \percent{68} scatter.
	The solid purple lines represent the mean luminosity functions from \Galform{} Monte Carlo realizations of each DM model and the corresponding shaded regions show their \percent{68} scatter. The green `corrected' Monte Carlo luminosity function is obtained by remapping the \MV{} of Monte Carlo satellite galaxies using the remapping relationships discussed in the text and shown in \figref{fig:Appendix:PCH-Tree_discrepancies:MV_remapping}.
	}%
	\label{fig:Appendix:PCH-Tree_discrepancies:LFs}%
	\vspace{-20pt}%
\end{figure}%

To attempt to address this overestimate, we performed the \myref{parkinson_generating_2008} calibration procedure using the \COCO{} simulations and found best-fitting values of $G_0{=}0.75$, $\gamma_1{=}0.1$ and $\gamma_2{=}-0.12$. The resulting Monte Carlo merger trees produce a better match with the \COCO{} merger trees; however, they remain discrepant across the range in satellite brightness. Consequently, when applying \Galform{} on the new Monte Carlo merger trees, this produces an overestimate of the faint end of the cumulative satellite galaxy luminosity function by a factor of ${\sim}1.6$ compared with that obtained by applying \Galform{} on the \COCO{} merger trees directly (cf. the \COCO{}+\Galform{} and Monte Carlo luminosity functions in \figref{fig:Appendix:PCH-Tree_discrepancies:LFs}). This discrepancy can be improved self-consistently only by altering the \myref{parkinson_generating_2008} algorithm, which would require more thorough investigation and possibly the introduction of one or more additional free parameters; this is beyond the scope of this work. 

\begin{figure}%
    \centering%
	\includegraphics[width=0.5\textwidth]{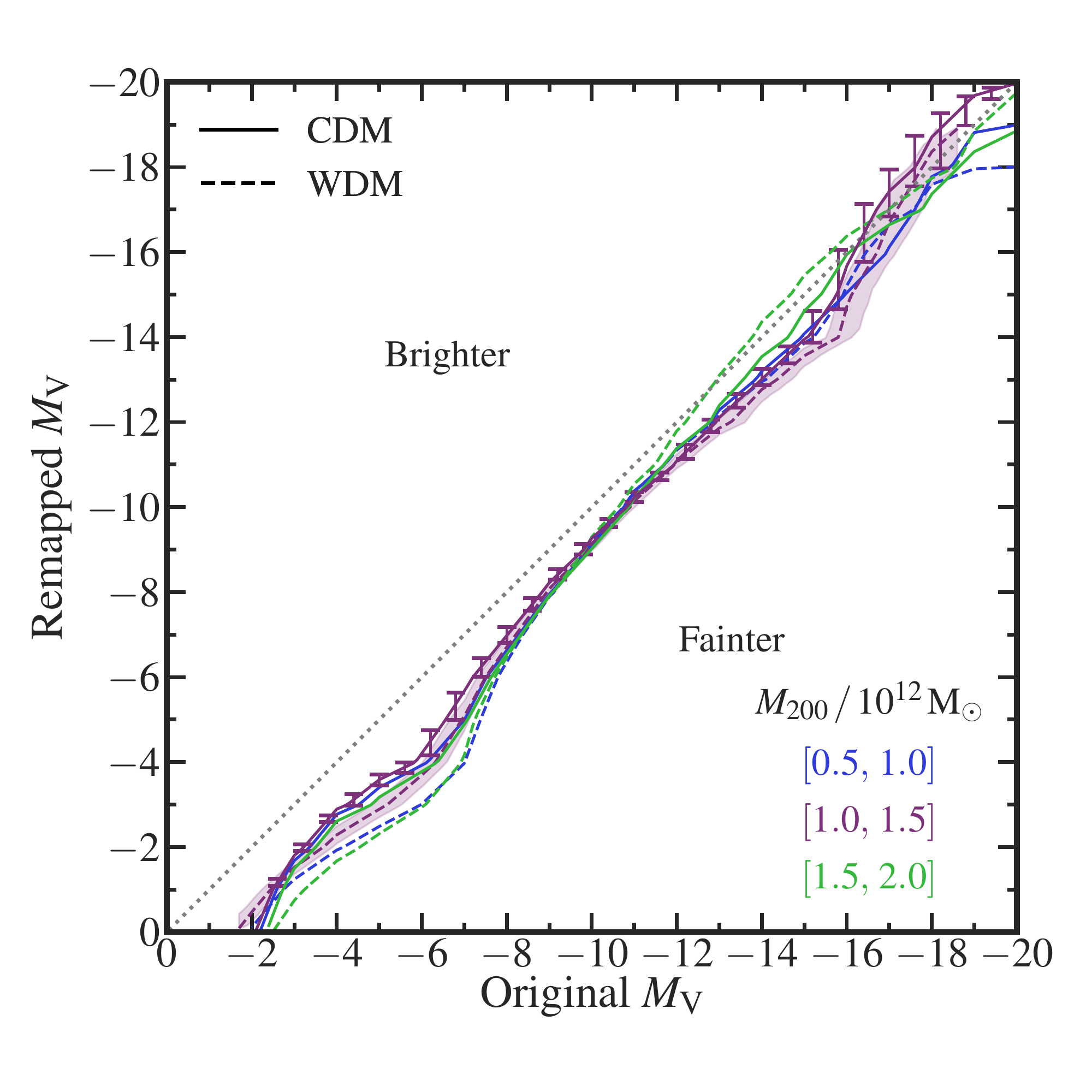}%
	\vspace{-17pt}%
	\caption{Functions to remap the \MV{} values of Monte Carlo \Galform{} satellite galaxies to new values that are consistent with the luminosity functions of \Galform{} applied to the \COCO{} suite. Only the functions for our fiducial \Galform{} model are shown, which are similar to the other parametrizations considered in this study. The dashed lines represent the remapping functions for the \keV[3.3] thermal relic \WDM{} model used in \wCOCO{}, and the solid lines show the functions for the \CDM{} model. In both cases, the lines are coloured by halo mass bin: ${\MNFW{}=\Msun[{\left[0.5,\, 1.0\right]\times10^{12}}]}$ (blue), \Msun[{\left[1.0,\, 1.5\right]\times10^{12}}] (purple) and \Msun[{\left[1.5,\, 2.0\right]\times10^{12}}] (green). The error bars (\CDM{}) and shaded region (\WDM{}) indicate the bootstrapped \percent{68} confidence intervals on the remapping relationships in the medium halo mass bin and are representative of the uncertainty in the other bins. The remapping functions are in excellent agreement across halo masses, apart from a small discrepancy at faint magnitudes.
	}%
	\label{fig:Appendix:PCH-Tree_discrepancies:MV_remapping}%
	\vspace{-20pt}%
\end{figure}

Instead, to obtain a satellite luminosity function for the Monte Carlo merger trees that is in agreement with the cosmological predictions, we map the satellite magnitude, \MV{}, predicted in the `Monte Carlo merger trees + \Galform{}' case to that of the `\COCO{} + \Galform{}' case by matching objects at fixed abundance, i.e. fixed \Nsat{} per host. By carrying out this procedure, we construct a remapping relationship between the `old' \MV{} and new values that are consistent with the \Nbody{} results. In \figref{fig:Appendix:PCH-Tree_discrepancies:MV_remapping}, we plot these relationships calculated for the \CDM{} and \keV[3.3] thermal relic \WDM{} models in three bins in halo mass. For clarity, we plot only the remapping functions obtained for our fiducial \Galform{} model with \zreion[7] and \vcut[30]. The error bars (\CDM{}) and shaded region (\WDM{}) indicate the bootstrapped \percent{68} confidence intervals on the remapping relationships in the halo mass bin ${\MNFW{}=\Msun[{\left[1,\, 1.5\right]\times10^{12}}]}$ and are representative of the uncertainties on the remapping functions in the other halo mass bins. 

The remapping functions are in excellent agreement across the range in halo mass in both DM models, and across almost the entire range in satellite brightness, although there is a small discrepancy between the \CDM{} and \WDM{} functions at the faint end. This corresponds to low-mass subhaloes near the cut-off scale in the \WDM{} power spectrum, whose properties differ the most from their equal mass \CDM{} counterparts. The differences in the formation histories of such objects in \WDM{} and \CDM{} models are modest \citep{lovell_properties_2014}, which explains the similarly modest discrepancy between the remapping functions of these models calculated here. We find similar results for the other parametrizations of reionization that we consider (not shown). Therefore, when calculating the results presented in \secref{sec:Galaxy_formation:Modelling_galform} we use the \CDM{} remapping relationships appropriate for each parametrization of reionization to adjust the \Galform{}-produced absolute magnitudes of dwarf satellite galaxies.

To check the remapping technique, we plot the corrected `Monte Carlo merger trees + \Galform{}' satellite luminosity function as a green curve in \figref{fig:Appendix:PCH-Tree_discrepancies:LFs}. By construction, the mean satellite count matches the \COCO{} predictions. More importantly, the \percent{68} scatter (represented by the green shaded region) is also in good agreement with the \Nbody{} results despite this not having been calibrated.

\section{Thermal relic mass constraints for different Galform results}
\label{app:Galform_model_results}

\begin{figure}%
    \centering%
	\includegraphics[width=\textwidth]{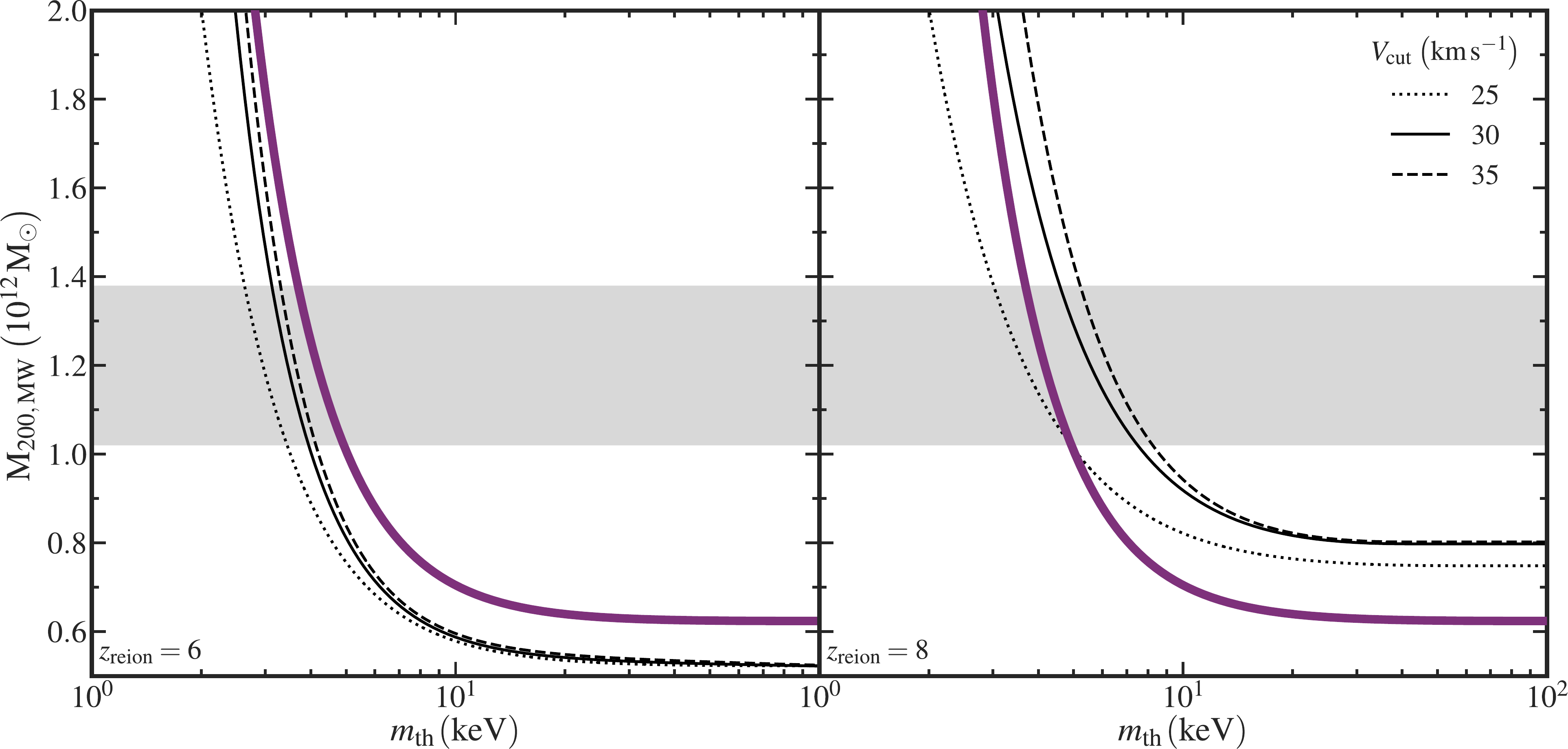}%
	\vspace{-10pt}%
	\caption{Constraints on the thermal relic particle mass obtained assuming three values of \vcut{} for \zreion[6] (left panel) and \zreion[8] (right panel) within the \Galform{} galaxy formation model. As in \figref{fig:Results:Fiducial_Galform_constraint}, parameter combinations to the left of and beneath the envelopes are ruled out with \percent{95} confidence. The thicker solid lines indicate the constraint envelope of our fiducial model with \zreion[7] and \vcut[30]. The shaded regions indicate the \percent{68} confidence interval on the mass of the MW halo from \citet{callingham_mass_2019}.
	}%
	\label{fig:Appendix:Galform_constraints:non_fid_z}%
	\vspace{-10pt}%
\end{figure}

Reionization plays an important role in the formation of low-mass dwarf galaxies and shapes the star formation history of the Universe more widely. In \Galform{}, reionization is described in terms of two key variables: the redshift by which reionization has ceased, \zreion{}, and the circular velocity cooling threshold, \vcut{}, below which galaxies and DM haloes are prevented after reionization from accreting cool gas from the intergalactic medium with which they might form more stars. To understand better how reionization affects the constraints on the thermal relic particle mass, we considered nine parameter combinations that span the allowed parameter range given our current observational constraints on reionization and galaxy formation models: $\vcut[{\left[25,\, 30,\, 35\right]}]$ and $\zreion[{\left[6,\,7,\, 8\right]}]$.  In the main paper, we showed how the DM particle mass constraints change when varying \vcut{} assuming \zreion[7], and when varying \zreion{} assuming \vcut[30], the results of which are presented in \figrefs{fig:Results:Fiducial_Galform_constraint}{fig:Results:Galform_fixed_vcut_zcut}. Here, we provide the constraints for parameter combinations assuming \zreion[6] and \zreion[8] (see \figref{fig:Appendix:Galform_constraints:non_fid_z}, left and right panels, respectively). In both cases, we also plot our fiducial constraint as a thicker solid line to facilitate easier comparison with these results. 

The dependence of the constraints on \zreion{} and \vcut{} demonstrated in \figref{fig:Results:Galform_fixed_vcut_zcut} also holds for the parameter choices  shown here. If reionization finishes later (\figref{fig:Appendix:Galform_constraints:non_fid_z} left panel), the strength of the constraints weakens considerably and the choice of \vcut{} becomes significantly more important. In \tabref{tab:Appendix:Galform_constraints:galform_model_constraints}, we provide the particle masses at and below which thermal relic \WDM{} models are excluded at \percent{95} confidence for each combination of reionization parameters that we consider in this study.

\begin{table}%
	\centering%
	\caption{Mass thresholds, \Mth{}, at and below which thermal relic models are excluded at \percent{95} confidence, for each \Galform{} model considered in this study.}%
	\label{tab:Appendix:Galform_constraints:galform_model_constraints}%
	\begin{tabularx}{\columnwidth}{YYYY}%
	    \hline
	    \vcut{} & \multicolumn{3}{c}{\zreion{}} \\
		& $6$ & $7$ & $8$ \\
		\hline
		\kms[25] & \keV[\zsixvtwofive{}] & \keV[\zsevenvtwofive{}] & \keV[\zeightvtwofive{}] \\
		\kms[30] & \keV[\zsixvthree{}] & \keV[\zsevenvthree{}] & \keV[\zeightvthree{}] \\
		\kms[35] & \keV[\zsixvthreefive{}] & \keV[\zsevenvthreefive{}] & \keV[\zeightvthreefive{}] \\
		\hline
	\end{tabularx}%
\end{table}%
\FloatBarrier

\acknowledgments

The authors thank the anonymous referee for valuable comments on the manuscript.
We also thank Piotr Ole\'{s}kiewicz and Andrew Griffin for their help to understand and modify the \Galform{} code, and Sownak Bose for his assistance with using \COCO{} simulation data.
ON thanks Steven Gillman for his hospitality during the final stages of this work.
ON was supported by the Science and Technology Facilities Council~(STFC) through grant ST/N50404X/1 and acknowledges support from the Institute for Computational Cosmology~(ICC) PhD Scholarships Fund and thanks the benefactors who fund it. ON also acknowledges financial support from the Project IDEXLYON at the University of Lyon under the Investments for the Future Programme (ANR-16-IDEX-0005) and supplementary financial support from La R\'{e}gion Auvergne-Rh\^{o}ne-Alpes. MC, ARJ and CSF were supported by STFC grant ST/L00075X/1. 
MC and CSF were supported by the ERC Advanced Investigator grant, DMIDAS [GA 786910].
MC acknowledges support by the EU Horizon 2020 research and innovation programme under a Marie Sk{\l}odowska-Curie grant agreement 794474 (DancingGalaxies).
This work used the DiRAC@Durham facility managed by the Institute for
Computational Cosmology on behalf of the STFC DiRAC HPC Facility~(\url{www.dirac.ac.uk}).
The equipment was funded by BEIS capital funding
via STFC capital grants ST/K00042X/1, ST/P002293/1, ST/R002371/1 and
ST/S002502/1, Durham University and STFC operations grant
ST/R000832/1. DiRAC is part of the National e-Infrastructure.

\textit{Software}: This research made use of \astropy{} \citep{the_astropy_collaboration_astropy_2013,the_astropy_collaboration_astropy_2018}, \matplotlib{} \citep{hunter_matplotlib_2007}, \numpy{} \citep{walt_numpy_2011,harris_array_2020}, \python{} \citep{rossum_python_1995,van_rossum_python_2009}, and \scipy{} \citep{jones_scipy_2011,virtanen_scipy_2020}. We thank their developers for maintaining them and making them freely available.

\section*{Data Availability}
The data used in this work are available upon reasonable request to the corresponding author.

\bibliographystyle{JHEP}
\bibliography{WDM_mass_constraints}

\providecommand{\href}[2]{#2}\begingroup\raggedright\begin{thebibliography}{100}

\bibitem{boyarsky_unidentified_2014}
A.~Boyarsky, O.~Ruchayskiy, D.~Iakubovskyi and J.~Franse, \emph{Unidentified
  {{Line}} in {{X}}-{{Ray Spectra}} of the {{Andromeda Galaxy}} and {{Perseus
  Galaxy Cluster}}},
  \href{https://doi.org/10.1103/PhysRevLett.113.251301}{\emph{Phys. Rev.}
  {\bfseries 113} (2014) L251301}.

\bibitem{bulbul_detection_2014}
E.~Bulbul, M.~Markevitch, A.~Foster, R.K.~Smith, M.~Loewenstein and
  S.W.~Randall, \emph{Detection of an {{Unidentified Emission Line}} in the
  {{Stacked X}}-{{Ray Spectrum}} of {{Galaxy Clusters}}},
  \href{https://doi.org/10.1088/0004-637X/789/1/13}{\emph{ApJ} {\bfseries 789}
  (2014) 13}.

\bibitem{boyarsky_checking_2015}
A.~Boyarsky, J.~Franse, D.~Iakubovskyi and O.~Ruchayskiy, \emph{Checking the
  dark matter origin of 3.53 {{keV}} line with the {{Milky Way}} center},
  \href{https://doi.org/10.1103/PhysRevLett.115.161301}{\emph{Phys. Rev.}
  {\bfseries 115} (2015) L161301}
  [\href{https://arxiv.org/abs/1408.2503}{{\ttfamily 1408.2503}}].

\bibitem{cappelluti_searching_2018}
N.~Cappelluti, E.~Bulbul, A.~Foster, P.~Natarajan, M.C.~Urry, M.W.~Bautz
  et~al., \emph{Searching for the 3.5 {{keV Line}} in the {{Deep Fields}} with
  {{Chandra}} : {{The}} 10 {{Ms Observations}}},
  \href{https://doi.org/10.3847/1538-4357/aaaa68}{\emph{ApJ} {\bfseries 854}
  (2018) 179}.

\bibitem{liu_current_2017}
J.~Liu, X.~Chen and X.~Ji, \emph{Current status of direct dark matter detection
  experiments}, \href{https://doi.org/10.1038/nphys4039}{\emph{Nature Physics}
  {\bfseries 13} (2017) 212}
  [\href{https://arxiv.org/abs/1709.00688}{{\ttfamily 1709.00688}}].

\bibitem{xenon_collaboration_first_2017}
{XENON Collaboration}, E.~Aprile, J.~Aalbers, F.~Agostini, M.~Alfonsi,
  F.D.~Amaro et~al., \emph{First {{Dark Matter Search Results}} from the
  {{XENON1T Experiment}}},
  \href{https://doi.org/10.1103/PhysRevLett.119.181301}{\emph{Phys. Rev.}
  {\bfseries 119} (2017) L181301}.

\bibitem{boehm_using_2014}
C.~B{\oe}hm, J.A.~Schewtschenko, R.J.~Wilkinson, C.M.~Baugh and S.~Pascoli,
  \emph{Using the {{Milky Way}} satellites to study interactions between cold
  dark matter and radiation},
  \href{https://doi.org/10.1093/mnrasl/slu115}{\emph{MNRAS} {\bfseries 445}
  (2014) L31}.

\bibitem{marsh_axion_2016}
D.J.E.~Marsh, \emph{Axion cosmology},
  \href{https://doi.org/10.1016/j.physrep.2016.06.005}{\emph{Phys. Rep.}
  {\bfseries 643} (2016) 1}.

\bibitem{escudero_fresh_2018}
M.~Escudero, L.~{Lopez-Honorez}, O.~Mena, S.~{Palomares-Ruiz} and
  P.~{Villanueva-Domingo}, \emph{A fresh look into the interacting dark matter
  scenario}, \href{https://doi.org/10.1088/1475-7516/2018/06/007}{\emph{JCAP}
  {\bfseries 2018} (2018) 007}.

\bibitem{sawala_apostle_2016}
T.~Sawala, C.S.~Frenk, A.~Fattahi, J.F.~Navarro, R.G.~Bower, R.A.~Crain et~al.,
  \emph{The {{APOSTLE}} simulations: Solutions to the {{Local Group}}'s cosmic
  puzzles}, \href{https://doi.org/10.1093/mnras/stw145}{\emph{MNRAS} {\bfseries
  457} (2016) 1931}.

\bibitem{avila-reese_formation_2001}
V.~{Avila-Reese}, P.~Col{\'i}n, O.~Valenzuela, E.~D'Onghia and C.~Firmani,
  \emph{Formation and {{Structure}} of {{Halos}} in a {{Warm Dark Matter
  Cosmology}}}, \href{https://doi.org/10.1086/322411}{\emph{ApJ} {\bfseries
  559} (2001) 516}.

\bibitem{bode_halo_2001}
P.~Bode, J.P.~Ostriker and N.~Turok, \emph{Halo {{Formation}} in {{Warm Dark
  Matter Models}}}, \href{https://doi.org/10.1086/321541}{\emph{ApJ} {\bfseries
  556} (2001) 93}.

\bibitem{shen_baryon_2014}
S.~Shen, P.~Madau, C.~Conroy, F.~Governato and L.~Mayer, \emph{The {{Baryon
  Cycle}} of {{Dwarf Galaxies}}: {{Dark}}, {{Bursty}}, {{Gas}}-rich
  {{Polluters}}}, \href{https://doi.org/10.1088/0004-637X/792/2/99}{\emph{ApJ}
  {\bfseries 792} (2014) 99}.

\bibitem{sawala_bent_2015}
T.~Sawala, C.S.~Frenk, A.~Fattahi, J.F.~Navarro, R.G.~Bower, R.A.~Crain et~al.,
  \emph{Bent by baryons: The low-mass galaxy-halo relation},
  \href{https://doi.org/10.1093/mnras/stu2753}{\emph{MNRAS} {\bfseries 448}
  (2015) 2941}.

\bibitem{sawala_chosen_2016}
T.~Sawala, C.S.~Frenk, A.~Fattahi, J.F.~Navarro, T.~Theuns, R.G.~Bower et~al.,
  \emph{The chosen few: The low-mass haloes that host faint galaxies},
  \href{https://doi.org/10.1093/mnras/stv2597}{\emph{MNRAS} {\bfseries 456}
  (2016) 85}.

\bibitem{wheeler_sweating_2015}
C.~Wheeler, J.~O{\~n}orbe, J.S.~Bullock, M.~{Boylan-Kolchin}, O.D.~Elbert,
  S.~{Garrison-Kimmel} et~al., \emph{Sweating the small stuff: Simulating dwarf
  galaxies, ultra-faint dwarf galaxies, and their own tiny satellites},
  \href{https://doi.org/10.1093/mnras/stv1691}{\emph{MNRAS} {\bfseries 453}
  (2015) 1305}.

\bibitem{koposov_luminosity_2008}
S.~Koposov, V.~Belokurov, N.W.~Evans, P.C.~Hewett, M.J.~Irwin, G.~Gilmore
  et~al., \emph{The {{Luminosity Function}} of the {{Milky Way Satellites}}},
  \href{https://doi.org/10.1086/589911}{\emph{ApJ} {\bfseries 686} (2008) 279}
  [\href{https://arxiv.org/abs/0706.2687}{{\ttfamily 0706.2687}}].

\bibitem{walsh_invisibles_2009}
S.M.~Walsh, B.~Willman and H.~Jerjen, \emph{The {{Invisibles A Detection
  Algorithm}} to {{Trace}} the {{Faintest Milky Way Satellites}}},
  \href{https://doi.org/10.1088/0004-6256/137/1/450}{\emph{AJ} {\bfseries 137}
  (2009) 450}.

\bibitem{hargis_too_2014}
J.R.~Hargis, B.~Willman and A.H.G.~Peter, \emph{Too {{Many}}, {{Too Few}}, or
  {{Just Right}}? {{The Predicted Number}} and {{Distribution}} of {{Milky Way
  Dwarf Galaxies}}},
  \href{https://doi.org/10.1088/2041-8205/795/1/L13}{\emph{ApJ} {\bfseries 795}
  (2014) L13}.

\bibitem{kennedy_constraining_2014}
R.~Kennedy, C.~Frenk, S.~Cole and A.~Benson, \emph{Constraining the warm dark
  matter particle mass with {{Milky Way}} satellites},
  \href{https://doi.org/10.1093/mnras/stu719}{\emph{MNRAS} {\bfseries 442}
  (2014) 2487}.

\bibitem{lovell_properties_2014}
M.R.~Lovell, C.S.~Frenk, V.R.~Eke, A.~Jenkins, L.~Gao and T.~Theuns, \emph{The
  properties of warm dark matter haloes},
  \href{https://doi.org/10.1093/mnras/stt2431}{\emph{MNRAS} {\bfseries 439}
  (2014) 300}.

\bibitem{newton_total_2018}
O.~Newton, M.~Cautun, A.~Jenkins, C.S.~Frenk and J.C.~Helly, \emph{The total
  satellite population of the {{Milky Way}}},
  \href{https://doi.org/10.1093/mnras/sty1085}{\emph{MNRAS} {\bfseries 479}
  (2018) 2853}.

\bibitem{nadler_modeling_2019}
E.O.~Nadler, Y.-Y.~Mao, G.M.~Green and R.H.~Wechsler, \emph{Modeling the
  {{Connection}} between {{Subhalos}} and {{Satellites}} in {{Milky
  Way}}\textendash like {{Systems}}},
  \href{https://doi.org/10.3847/1538-4357/ab040e}{\emph{ApJ} {\bfseries 873}
  (2019) 34}.

\bibitem{springel_aquarius_2008}
V.~Springel, J.~Wang, M.~Vogelsberger, A.~Ludlow, A.~Jenkins, A.~Helmi et~al.,
  \emph{The {{Aquarius Project}}: The subhaloes of galactic haloes},
  \href{https://doi.org/10.1111/j.1365-2966.2008.14066.x}{\emph{MNRAS}
  {\bfseries 391} (2008) 1685}.

\bibitem{onions_subhaloes_2012}
J.~Onions, A.~Knebe, F.R.~Pearce, S.I.~Muldrew, H.~Lux, S.R.~Knollmann et~al.,
  \emph{Subhaloes going {{Notts}}: The subhalo-finder comparison project},
  \href{https://doi.org/10.1111/j.1365-2966.2012.20947.x}{\emph{MNRAS}
  {\bfseries 423} (2012) 1200}.

\bibitem{van_den_bosch_dark_2018}
F.C.~{van den Bosch} and G.~Ogiya, \emph{Dark matter substructure in numerical
  simulations: A tale of discreteness noise, runaway instabilities, and
  artificial disruption},
  \href{https://doi.org/10.1093/mnras/sty084}{\emph{MNRAS} {\bfseries 475}
  (2018) 4066}.

\bibitem{kaviraj_tidal_2012}
S.~Kaviraj, D.~Darg, C.~Lintott, K.~Schawinski and J.~Silk, \emph{Tidal dwarf
  galaxies in the nearby {{Universe}}},
  \href{https://doi.org/10.1111/j.1365-2966.2011.19673.x}{\emph{MNRAS}
  {\bfseries 419} (2012) 70}.

\bibitem{lisenfeld_molecular_2016}
U.~Lisenfeld, J.~Braine, P.A.~Duc, M.~Boquien, E.~Brinks, F.~Bournaud et~al.,
  \emph{Molecular gas and star formation in the tidal dwarf galaxy {{VCC}}
  2062}, \href{https://doi.org/10.1051/0004-6361/201527887}{\emph{A\&A}
  {\bfseries 590} (2016) A92}.

\bibitem{ploeckinger_tidal_2018}
S.~Ploeckinger, K.~Sharma, J.~Schaye, R.A.~Crain, M.~Schaller and C.~Barber,
  \emph{Tidal dwarf galaxies in cosmological simulations},
  \href{https://doi.org/10.1093/mnras/stx2787}{\emph{MNRAS} {\bfseries 474}
  (2018) 580}.

\bibitem{haslbauer_galaxies_2019}
M.~Haslbauer, J.~Dabringhausen, P.~Kroupa, B.~Javanmardi and I.~Banik,
  \emph{Galaxies lacking dark matter in the {{Illustris}} simulation},
  \href{https://doi.org/10.1051/0004-6361/201833771}{\emph{A\&A} {\bfseries
  626} (2019) A47}.

\bibitem{frenk_formation_1988}
C.S.~Frenk, S.D.M.~White, M.~Davis and G.~Efstathiou, \emph{The formation of
  dark halos in a universe dominated by cold dark matter},
  \href{https://doi.org/10.1086/166213}{\emph{ApJ} {\bfseries 327} (1988) 507}.

\bibitem{hellwing_copernicus_2016}
W.A.~Hellwing, C.S.~Frenk, M.~Cautun, S.~Bose, J.~Helly, A.~Jenkins et~al.,
  \emph{The {{Copernicus Complexio}}: A high-resolution view of the small-scale
  {{Universe}}}, \href{https://doi.org/10.1093/mnras/stw214}{\emph{MNRAS}
  {\bfseries 457} (2016) 3492}.

\bibitem{bose_copernicus_2016}
S.~Bose, W.A.~Hellwing, C.S.~Frenk, A.~Jenkins, M.R.~Lovell, J.C.~Helly et~al.,
  \emph{The {{Copernicus Complexio}}: Statistical properties of warm dark
  matter haloes}, \href{https://doi.org/10.1093/mnras/stv2294}{\emph{MNRAS}
  {\bfseries 455} (2016) 318}.

\bibitem{komatsu_seven-year_2011}
E.~Komatsu, K.M.~Smith, J.~Dunkley, C.L.~Bennett, B.~Gold, G.~Hinshaw et~al.,
  \emph{Seven-year {{Wilkinson Microwave Anisotropy Probe}} ({{WMAP}})
  {{Observations}}: {{Cosmological Interpretation}}},
  \href{https://doi.org/10.1088/0067-0049/192/2/18}{\emph{ApJS} {\bfseries 192}
  (2011) 18}.

\bibitem{wang_discreteness_2007}
J.~Wang and S.D.M.~White, \emph{Discreteness effects in simulations of hot/warm
  dark matter},
  \href{https://doi.org/10.1111/j.1365-2966.2007.12053.x}{\emph{MNRAS}
  {\bfseries 380} (2007) 93}.

\bibitem{angulo_warm_2013}
R.E.~Angulo, O.~Hahn and T.~Abel, \emph{The warm dark matter halo mass function
  below the cut-off scale},
  \href{https://doi.org/10.1093/mnras/stt1246}{\emph{MNRAS} {\bfseries 434}
  (2013) 3337}.

\bibitem{green_tidal_2019}
S.B.~Green and F.C.~{van den Bosch}, \emph{The tidal evolution of dark matter
  substructure \textendash{} {{I}}. subhalo density profiles},
  \href{https://doi.org/10.1093/mnras/stz2767}{\emph{MNRAS} {\bfseries 490}
  (2019) 2091}.

\bibitem{errani_asymptotic_2021}
R.~Errani and J.F.~Navarro, \emph{The asymptotic tidal remnants of cold dark
  matter subhaloes}, \href{https://doi.org/10.1093/mnras/stab1215}{\emph{MNRAS}
  {\bfseries 505} (2021) 18}
  [\href{https://arxiv.org/abs/2011.07077}{{\ttfamily 2011.07077}}].

\bibitem{green_tidal_2021}
S.B.~Green, F.C.~{van den Bosch} and F.~Jiang, \emph{The tidal evolution of
  dark matter substructure \textendash{} {{II}}. {{The}} impact of artificial
  disruption on subhalo mass functions and radial profiles},
  \href{https://doi.org/10.1093/mnras/stab696}{\emph{MNRAS} {\bfseries 503}
  (2021) 4075} [\href{https://arxiv.org/abs/2103.01227}{{\ttfamily
  2103.01227}}].

\bibitem{tollerud_hundreds_2008}
E.J.~Tollerud, J.S.~Bullock, L.E.~Strigari and B.~Willman, \emph{Hundreds of
  {{Milky Way Satellites}}? {{Luminosity Bias}} in the {{Satellite Luminosity
  Function}}}, \href{https://doi.org/10.1086/592102}{\emph{ApJ} {\bfseries 688}
  (2008) 277}.

\bibitem{alam_eleventh_2015}
S.~Alam, F.D.~Albareti, C.A.~Prieto, F.~Anders, S.F.~Anderson, {Timothy
  Anderton} et~al., \emph{The {{Eleventh}} and {{Twelfth Data Releases}} of the
  {{Sloan Digital Sky Survey}}: {{Final Data}} from {{SDSS}}-{{III}}},
  \href{https://doi.org/10.1088/0067-0049/219/1/12}{\emph{ApJS} {\bfseries 219}
  (2015) 12}.

\bibitem{bechtol_eight_2015}
K.~Bechtol, A.~{Drlica-Wagner}, E.~Balbinot, A.~Pieres, J.D.~Simon, B.~Yanny
  et~al., \emph{Eight {{New Milky Way Companions Discovered}} in {{First}}-year
  {{Dark Energy Survey Data}}},
  \href{https://doi.org/10.1088/0004-637X/807/1/50}{\emph{ApJ} {\bfseries 807}
  (2015) 50}.

\bibitem{drlica-wagner_eight_2015}
A.~{Drlica-Wagner}, K.~Bechtol, E.S.~Rykoff, E.~Luque, A.~Queiroz, Y.-Y.~Mao
  et~al., \emph{Eight {{Ultra}}-faint {{Galaxy Candidates Discovered}} in
  {{Year Two}} of the {{Dark Energy Survey}}},
  \href{https://doi.org/10.1088/0004-637X/813/2/109}{\emph{ApJ} {\bfseries 813}
  (2015) 109}.

\bibitem{bose_no_2019}
S.~Bose, C.S.~Frenk, A.~Jenkins, A.~Fattahi, F.A.~G{\'o}mez, R.J.J.~Grand
  et~al., \emph{No cores in dark matter-dominated dwarf galaxies with bursty
  star formation histories},
  \href{https://doi.org/10.1093/mnras/stz1168}{\emph{MNRAS} {\bfseries 486}
  (2019) 4790}.

\bibitem{newton_mw_2018}
O.~Newton and M.~Cautun, ``{{MW Satellite LF}}: V1.0.0 release.'' Zenodo, Mar.,
  2018.
\newblock 10.5281/zenodo.1205622.

\bibitem{watkins_substructure_2009}
L.L.~Watkins, N.W.~Evans, V.~Belokurov, M.C.~Smith, P.C.~Hewett, D.M.~Bramich
  et~al., \emph{Substructure revealed by {{RR Lyraes}} in {{SDSS Stripe}} 82},
  \href{https://doi.org/10.1111/j.1365-2966.2009.15242.x}{\emph{MNRAS}
  {\bfseries 398} (2009) 1757}.

\bibitem{mcconnachie_observed_2012}
A.W.~McConnachie, \emph{The {{Observed Properties}} of {{Dwarf Galaxies}} in
  and around the {{Local Group}}},
  \href{https://doi.org/10.1088/0004-6256/144/1/4}{\emph{AJ} {\bfseries 144}
  (2012) 4}.

\bibitem{kim_heros_2015}
D.~Kim, H.~Jerjen, D.~Mackey, G.S.D.~Costa and A.P.~Milone, \emph{A {{Hero}}'s
  {{Dark Horse}}: {{Discovery}} of an {{Ultra}}-faint {{Milky Way Satellite}}
  in {{Pegasus}}},
  \href{https://doi.org/10.1088/2041-8205/804/2/L44}{\emph{ApJ} {\bfseries 804}
  (2015) L44}.

\bibitem{koposov_kinematics_2015}
S.E.~Koposov, A.R.~Casey, V.~Belokurov, J.R.~Lewis, G.~Gilmore, C.~Worley
  et~al., \emph{Kinematics and {{Chemistry}} of {{Recently Discovered
  Reticulum}} 2 and {{Horologium}} 1 {{Dwarf Galaxies}}},
  \href{https://doi.org/10.1088/0004-637X/811/1/62}{\emph{ApJ} {\bfseries 811}
  (2015) 62}.

\bibitem{jethwa_magellanic_2016}
P.~Jethwa, D.~Erkal and V.~Belokurov, \emph{A {{Magellanic}} origin of the
  {{DES}} dwarfs}, \href{https://doi.org/10.1093/mnras/stw1343}{\emph{MNRAS}
  {\bfseries 461} (2016) 2212}.

\bibitem{kim_portrait_2016}
D.~Kim, H.~Jerjen, M.~Geha, A.~Chiti, A.P.~Milone, G.D.~Costa et~al.,
  \emph{Portrait of a {{Dark Horse}}: A {{Photometric}} and {{Spectroscopic
  Study}} of the {{Ultra}}-faint {{Milky Way Satellite Pegasus III}}},
  \href{https://doi.org/10.3847/0004-637X/833/1/16}{\emph{ApJ} {\bfseries 833}
  (2016) 16}.

\bibitem{walker_magellan/m2fs_2016}
M.G.~Walker, M.~Mateo, E.W.~Olszewski, S.~Koposov, V.~Belokurov, {Prashin
  Jethwa} et~al., \emph{Magellan/{{M2FS Spectroscopy}} of {{Tucana}} 2 and
  {{Grus}} 1}, \href{https://doi.org/10.3847/0004-637X/819/1/53}{\emph{ApJ}
  {\bfseries 819} (2016) 53}.

\bibitem{carlin_deep_2017}
J.L.~Carlin, D.J.~Sand, R.R.~Mu{\~n}oz, K.~Spekkens, B.~Willman, {Denija
  Crnojevi\'c} et~al., \emph{Deep {{Subaru Hyper Suprime}}-{{Cam Observations}}
  of {{Milky Way Satellites Columba I}} and {{Triangulum II}}},
  \href{https://doi.org/10.3847/1538-3881/aa94d0}{\emph{AJ} {\bfseries 154}
  (2017) 267}.

\bibitem{li_farthest_2017}
T.S.~Li, J.D.~Simon, A.~{Drlica-Wagner}, K.~Bechtol, M.Y.~Wang,
  J.~{Garc{\'i}a-Bellido} et~al., \emph{Farthest {{Neighbor}}: {{The Distant
  Milky Way Satellite Eridanus II}}},
  \href{https://doi.org/10.3847/1538-4357/aa6113}{\emph{ApJ} {\bfseries 838}
  (2017) 8}.

\bibitem{drlica-wagner_milky_2020}
A.~{Drlica-Wagner}, K.~Bechtol, S.~Mau, M.~McNanna, E.O.~Nadler, A.B.~Pace
  et~al., \emph{Milky {{Way Satellite Census}}. {{I}}. {{The Observational
  Selection Function}} for {{Milky Way Satellites}} in {{DES Y3}} and
  {{Pan}}-{{STARRS DR1}}},
  \href{https://doi.org/10.3847/1538-4357/ab7eb9}{\emph{ApJ} {\bfseries 893}
  (2020) 47}.

\bibitem{press_formation_1974}
W.H.~Press and P.~Schechter, \emph{Formation of {{Galaxies}} and {{Clusters}}
  of {{Galaxies}} by {{Self}}-{{Similar Gravitational Condensation}}},
  \href{https://doi.org/10.1086/152650}{\emph{ApJ} {\bfseries 187} (1974) 425}.

\bibitem{bond_excursion_1991}
J.R.~Bond, S.~Cole, G.~Efstathiou and N.~Kaiser, \emph{Excursion set mass
  functions for hierarchical {{Gaussian}} fluctuations},
  \href{https://doi.org/10.1086/170520}{\emph{ApJ} {\bfseries 379} (1991) 440}.

\bibitem{bower_evolution_1991}
R.G.~Bower, \emph{The evolution of groups of galaxies in the
  {{Press}}\textendash{{Schechter}} formalism},
  \href{https://doi.org/10.1093/mnras/248.2.332}{\emph{MNRAS} {\bfseries 248}
  (1991) 332}.

\bibitem{lacey_merger_1993}
C.~Lacey and S.~Cole, \emph{Merger rates in hierarchical models of galaxy
  formation}, \href{https://doi.org/10.1093/mnras/262.3.627}{\emph{MNRAS}
  {\bfseries 262} (1993) 627}.

\bibitem{parkinson_generating_2008}
H.~Parkinson, S.~Cole and J.~Helly, \emph{Generating dark matter halo merger
  trees}, \href{https://doi.org/10.1111/j.1365-2966.2007.12517.x}{\emph{MNRAS}
  {\bfseries 383} (2008) 557}.

\bibitem{benson_dark_2013}
A.J.~Benson, A.~Farahi, S.~Cole, L.A.~Moustakas, A.~Jenkins, M.~Lovell et~al.,
  \emph{Dark matter halo merger histories beyond cold dark matter \textendash{}
  {{I}}. {{Methods}} and application to warm dark matter},
  \href{https://doi.org/10.1093/mnras/sts159}{\emph{MNRAS} {\bfseries 428}
  (2013) 1774}.

\bibitem{lovell_satellite_2016}
M.R.~Lovell, S.~Bose, A.~Boyarsky, S.~Cole, C.S.~Frenk, V.~{Gonzalez-Perez}
  et~al., \emph{Satellite galaxies in semi-analytic models of galaxy formation
  with sterile neutrino dark matter},
  \href{https://doi.org/10.1093/mnras/stw1317}{\emph{MNRAS} {\bfseries 461}
  (2016) 60}.

\bibitem{leo_new_2018}
M.~Leo, C.M.~Baugh, B.~Li and S.~Pascoli, \emph{A new smooth-k space filter
  approach to calculate halo abundances},
  \href{https://doi.org/10.1088/1475-7516/2018/04/010}{\emph{JCAP} {\bfseries
  2018} (2018) 010}.

\bibitem{schneider_halo_2013}
A.~Schneider, R.E.~Smith and D.~Reed, \emph{Halo mass function and the free
  streaming scale}, \href{https://doi.org/10.1093/mnras/stt829}{\emph{MNRAS}
  {\bfseries 433} (2013) 1573}.

\bibitem{giocoli_analytical_2008}
C.~Giocoli, L.~Pieri and G.~Tormen, \emph{Analytical approach to subhalo
  population in dark matter haloes},
  \href{https://doi.org/10.1111/j.1365-2966.2008.13283.x}{\emph{MNRAS}
  {\bfseries 387} (2008) 689}.

\bibitem{schneider_structure_2015}
A.~Schneider, \emph{Structure formation with suppressed small-scale
  perturbations}, \href{https://doi.org/10.1093/mnras/stv1169}{\emph{MNRAS}
  {\bfseries 451} (2015) 3117}.

\bibitem{viel_constraining_2005}
M.~Viel, J.~Lesgourgues, M.G.~Haehnelt, S.~Matarrese and A.~Riotto,
  \emph{Constraining warm dark matter candidates including sterile neutrinos
  and light gravitinos with {{WMAP}} and the {{Lyman}}-\$\textbackslash
  ensuremath\{\textbackslash alpha\}\$ forest},
  \href{https://doi.org/10.1103/PhysRevD.71.063534}{\emph{Phys. Rev. D}
  {\bfseries 71} (2005) 063534}.

\bibitem{springel_populating_2001}
V.~Springel, S.D.M.~White, G.~Tormen and G.~Kauffmann, \emph{Populating a
  cluster of galaxies \textendash{} {{I}}. {{Results}} at z = 0},
  \href{https://doi.org/10.1046/j.1365-8711.2001.04912.x}{\emph{MNRAS}
  {\bfseries 328} (2001) 726}.

\bibitem{dolag_substructures_2009}
K.~Dolag, S.~Borgani, G.~Murante and V.~Springel, \emph{Substructures in
  hydrodynamical cluster simulations},
  \href{https://doi.org/10.1111/j.1365-2966.2009.15034.x}{\emph{MNRAS}
  {\bfseries 399} (2009) 497}.

\bibitem{boylan-kolchin_theres_2010}
M.~{Boylan-Kolchin}, V.~Springel, S.D.M.~White and A.~Jenkins, \emph{There's no
  place like home? {{Statistics}} of {{Milky Way}}-mass dark matter haloes},
  \href{https://doi.org/10.1111/j.1365-2966.2010.16774.x}{\emph{MNRAS}
  {\bfseries 406} (2010) 896}.

\bibitem{cautun_subhalo_2014}
M.~Cautun, W.A.~Hellwing, R.~{van de Weygaert}, C.S.~Frenk, B.J.T.~Jones and
  T.~Sawala, \emph{Subhalo statistics of galactic haloes: Beyond the resolution
  limit}, \href{https://doi.org/10.1093/mnras/stu1829}{\emph{MNRAS} {\bfseries
  445} (2014) 1820}.

\bibitem{wang_missing_2012}
J.~Wang, C.S.~Frenk, J.F.~Navarro, L.~Gao and T.~Sawala, \emph{The missing
  massive satellites of the {{Milky Way}}},
  \href{https://doi.org/10.1111/j.1365-2966.2012.21357.x}{\emph{MNRAS}
  {\bfseries 424} (2012) 2715}.

\bibitem{wang_mass_2020}
W.~Wang, J.~Han, M.~Cautun, Z.~Li and M.N.~Ishigaki, \emph{The mass of our
  {{Milky Way}}}, \href{https://doi.org/10.1007/s11433-019-1541-6}{\emph{Sci.
  China Phys. Mech. Astron.} {\bfseries 63} (2020) 109801}.

\bibitem{polisensky_constraints_2011}
E.~Polisensky and M.~Ricotti, \emph{Constraints on the dark matter particle
  mass from the number of {{Milky Way}} satellites},
  \href{https://doi.org/10.1103/PhysRevD.83.043506}{\emph{Phys. Rev. D}
  {\bfseries 83} (2011) 043506}.

\bibitem{callingham_mass_2019}
T.M.~Callingham, M.~Cautun, A.J.~Deason, C.S.~Frenk, W.~Wang, F.A.~G{\'o}mez
  et~al., \emph{The mass of the {{Milky Way}} from satellite dynamics},
  \href{https://doi.org/10.1093/mnras/stz365}{\emph{MNRAS} {\bfseries 484}
  (2019) 5453}.

\bibitem{cautun_milky_2020}
M.~Cautun, A.~{Ben{\'i}tez-Llambay}, A.J.~Deason, C.S.~Frenk, A.~Fattahi,
  F.A.~G{\'o}mez et~al., \emph{The milky way total mass profile as inferred
  from {{Gaia DR2}}},
  \href{https://doi.org/10.1093/mnras/staa1017}{\emph{MNRAS} {\bfseries 494}
  (2020) 4291}.

\bibitem{cole_recipe_1994}
S.~Cole, A.~{Arag{\'o}n-Salamanca}, C.S.~Frenk, J.F.~Navarro and S.E.~Zepf,
  \emph{A recipe for galaxy formation},
  \href{https://doi.org/10.1093/mnras/271.4.781}{\emph{MNRAS} {\bfseries 271}
  (1994) 781}.

\bibitem{cole_hierarchical_2000}
S.~Cole, C.G.~Lacey, C.M.~Baugh and C.S.~Frenk, \emph{Hierarchical galaxy
  formation},
  \href{https://doi.org/10.1046/j.1365-8711.2000.03879.x}{\emph{MNRAS}
  {\bfseries 319} (2000) 168}.

\bibitem{lacey_unified_2016}
C.G.~Lacey, C.M.~Baugh, C.S.~Frenk, A.J.~Benson, R.G.~Bower, S.~Cole et~al.,
  \emph{A unified multiwavelength model of galaxy formation},
  \href{https://doi.org/10.1093/mnras/stw1888}{\emph{MNRAS} {\bfseries 462}
  (2016) 3854}.

\bibitem{benson_effects_2002}
A.J.~Benson, C.G.~Lacey, C.M.~Baugh, S.~Cole and C.S.~Frenk, \emph{The effects
  of photoionization on galaxy formation \textendash{} {{I}}. {{Model}} and
  results at z=0},
  \href{https://doi.org/10.1046/j.1365-8711.2002.05387.x}{\emph{MNRAS}
  {\bfseries 333} (2002) 156}.

\bibitem{font_population_2011}
A.S.~Font, A.J.~Benson, R.G.~Bower, C.S.~Frenk, A.~Cooper, G.~DeLucia et~al.,
  \emph{The population of {{Milky Way}} satellites in the {{$\Lambda$}} cold
  dark matter cosmology},
  \href{https://doi.org/10.1111/j.1365-2966.2011.19339.x}{\emph{MNRAS}
  {\bfseries 417} (2011) 1260}.

\bibitem{bose_imprint_2018}
S.~Bose, A.J.~Deason and C.S.~Frenk, \emph{The {{Imprint}} of {{Cosmic
  Reionization}} on the {{Luminosity Function}} of {{Galaxies}}},
  \href{https://doi.org/10.3847/1538-4357/aacbc4}{\emph{ApJ} {\bfseries 863}
  (2018) 123}.

\bibitem{jethwa_upper_2017}
P.~Jethwa, D.~Erkal and V.~Belokurov, \emph{The upper bound on the lowest mass
  halo}, \href{https://doi.org/10.1093/mnras/stx2330}{\emph{MNRAS} {\bfseries
  473} (2017) 2060}.

\bibitem{mason_universe_2018}
C.A.~Mason, T.~Treu, M.~Dijkstra, A.~Mesinger, M.~Trenti, L.~Pentericci et~al.,
  \emph{The {{Universe Is Reionizing}} at z \$\textbackslash sim\$ 7:
  {{Bayesian Inference}} of the {{IGM Neutral Fraction Using
  Ly}}\$\textbackslash upalpha\$ {{Emission}} from {{Galaxies}}},
  \href{https://doi.org/10.3847/1538-4357/aab0a7}{\emph{ApJ} {\bfseries 856}
  (2018) 2}.

\bibitem{planck_collaboration_planck_2020}
{Planck Collaboration}, N.~Aghanim, Y.~Akrami, M.~Ashdown, J.~Aumont,
  C.~Baccigalupi et~al., \emph{Planck 2018 results - {{VI}}. {{Cosmological}}
  parameters}, \href{https://doi.org/10.1051/0004-6361/201833910}{\emph{A\&A}
  {\bfseries 641} (2020) A6}
  [\href{https://arxiv.org/abs/1807.06209}{{\ttfamily 1807.06209}}].

\bibitem{okamoto_mass_2008}
T.~Okamoto, L.~Gao and T.~Theuns, \emph{Mass loss of galaxies due to an
  ultraviolet background},
  \href{https://doi.org/10.1111/j.1365-2966.2008.13830.x}{\emph{MNRAS}
  {\bfseries 390} (2008) 920}.

\bibitem{robertson_cosmic_2015}
B.E.~Robertson, R.S.~Ellis, S.R.~Furlanetto and J.S.~Dunlop, \emph{{{COSMIC
  REIONIZATION AND EARLY STAR}}-{{FORMING GALAXIES}}: {{A JOINT ANALYSIS OF NEW
  CONSTRAINTS FROM PLANCK AND THE HUBBLE SPACE TELESCOPE}}},
  \href{https://doi.org/10.1088/2041-8205/802/2/L19}{\emph{ApJ} {\bfseries 802}
  (2015) L19}.

\bibitem{banados_800-million-solar-mass_2018}
E.~Ba{\~n}ados, B.P.~Venemans, C.~Mazzucchelli, E.P.~Farina, F.~Walter, F.~Wang
  et~al., \emph{An 800-million-solar-mass black hole in a significantly neutral
  {{Universe}} at a redshift of 7.5},
  \href{https://doi.org/10.1038/nature25180}{\emph{Nature} {\bfseries 553}
  (2018) 473}.

\bibitem{davies_quantitative_2018}
F.B.~Davies, J.F.~Hennawi, E.~Ba{\~n}ados, Z.~Luki{\'c}, R.~Decarli, X.~Fan
  et~al., \emph{Quantitative {{Constraints}} on the {{Reionization History}}
  from the {{IGM Damping Wing Signature}} in {{Two Quasars}} at z
  \$\textbackslash greater\$ 7},
  \href{https://doi.org/10.3847/1538-4357/aad6dc}{\emph{ApJ} {\bfseries 864}
  (2018) 142}.

\bibitem{nadler_milky_2020}
E.O.~Nadler, R.H.~Wechsler, K.~Bechtol, Y.-Y.~Mao, G.~Green, A.~{Drlica-Wagner}
  et~al., \emph{Milky {{Way Satellite Census}}. {{II}}.
  {{Galaxy}}\textendash{{Halo Connection Constraints Including}} the {{Impact}}
  of the {{Large Magellanic Cloud}}},
  \href{https://doi.org/10.3847/1538-4357/ab846a}{\emph{ApJ} {\bfseries 893}
  (2020) 48} [\href{https://arxiv.org/abs/1912.03303}{{\ttfamily 1912.03303}}].

\bibitem{safarzadeh_limit_2018}
M.~Safarzadeh, E.~Scannapieco and A.~Babul, \emph{A {{Limit}} on the {{Warm
  Dark Matter Particle Mass}} from the {{Redshifted}} 21 cm {{Absorption
  Line}}}, \href{https://doi.org/10.3847/2041-8213/aac5e0}{\emph{ApJ}
  {\bfseries 859} (2018) L18}.

\bibitem{baur_lyman-alpha_2016}
J.~Baur, N.~{Palanque-Delabrouille}, C.~Y{\`e}che, C.~Magneville and M.~Viel,
  \emph{Lyman-alpha forests cool warm dark matter},
  \href{https://doi.org/10.1088/1475-7516/2016/08/012}{\emph{JCAP} {\bfseries
  2016} (2016) 012}.

\bibitem{viel_warm_2013}
M.~Viel, G.D.~Becker, J.S.~Bolton and M.G.~Haehnelt, \emph{Warm dark matter as
  a solution to the small scale crisis: {{New}} constraints from high redshift
  {{Lyman}}-\$\textbackslash ensuremath\{\textbackslash alpha\}\$ forest data},
  \href{https://doi.org/10.1103/PhysRevD.88.043502}{\emph{Phys. Rev. D}
  {\bfseries 88} (2013) 043502}.

\bibitem{irsic_new_2017}
V.~Ir{\v s}i{\v c}, M.~Viel, M.G.~Haehnelt, J.S.~Bolton, S.~Cristiani,
  G.D.~Becker et~al., \emph{New constraints on the free-streaming of warm dark
  matter from intermediate and small scale {{Lyman}}-\$\textbackslash
  ensuremath\{\textbackslash alpha\}\$ forest data},
  \href{https://doi.org/10.1103/PhysRevD.96.023522}{\emph{Phys. Rev. D}
  {\bfseries 96} (2017) 023522}.

\bibitem{hsueh_sharp_2020}
J.-W.~Hsueh, W.~Enzi, S.~Vegetti, M.W.~Auger, C.D.~Fassnacht, G.~Despali
  et~al., \emph{{{SHARP}} \textendash{} {{VII}}. {{New}} constraints on the
  dark matter free-streaming properties and substructure abundance from
  gravitationally lensed quasars},
  \href{https://doi.org/10.1093/mnras/stz3177}{\emph{MNRAS} {\bfseries 492}
  (2020) 3047} [\href{https://arxiv.org/abs/1905.04182}{{\ttfamily
  1905.04182}}].

\bibitem{gnedin_tidal_1999}
O.Y.~Gnedin, L.~Hernquist and J.P.~Ostriker, \emph{Tidal {{Shocking}} by
  {{Extended Mass Distributions}}},
  \href{https://doi.org/10.1086/306910}{\emph{ApJ} {\bfseries 514} (1999) 109}.

\bibitem{brooks_why_2014}
A.M.~Brooks and A.~Zolotov, \emph{Why {{Baryons Matter}}: {{The Kinematics}} of
  {{Dwarf Spheroidal Satellites}}},
  \href{https://doi.org/10.1088/0004-637X/786/2/87}{\emph{ApJ} {\bfseries 786}
  (2014) 87}.

\bibitem{garrison-kimmel_not_2017}
S.~{Garrison-Kimmel}, A.~Wetzel, J.S.~Bullock, P.F.~Hopkins,
  M.~{Boylan-Kolchin}, C.-A.~{Faucher-Gigu{\`e}re} et~al., \emph{Not so lumpy
  after all: Modelling the depletion of dark matter subhaloes by {{Milky
  Way}}-like galaxies},
  \href{https://doi.org/10.1093/mnras/stx1710}{\emph{MNRAS} {\bfseries 471}
  (2017) 1709}.

\bibitem{sawala_shaken_2017}
T.~Sawala, P.~Pihajoki, P.H.~Johansson, C.S.~Frenk, J.F.~Navarro, K.A.~Oman
  et~al., \emph{Shaken and stirred: The {{Milky Way}}'s dark substructures},
  \href{https://doi.org/10.1093/mnras/stx360}{\emph{MNRAS} {\bfseries 467}
  (2017) 4383}.

\bibitem{richings_subhalo_2020}
J.~Richings, C.~Frenk, A.~Jenkins, A.~Robertson, A.~Fattahi, R.J.J.~Grand
  et~al., \emph{Subhalo destruction in the {{Apostle}} and {{Auriga}}
  simulations}, \href{https://doi.org/10.1093/mnras/stz3448}{\emph{MNRAS}
  {\bfseries 492} (2020) 5780}.

\bibitem{richings_high-resolution_2021}
J.~Richings, C.~Frenk, A.~Jenkins, A.~Robertson and M.~Schaller, \emph{A
  high-resolution cosmological simulation of a strong gravitational lens},
  \href{https://doi.org/10.1093/mnras/staa4013}{\emph{MNRAS} {\bfseries 501}
  (2021) 4657} [\href{https://arxiv.org/abs/2005.14495}{{\ttfamily
  2005.14495}}].

\bibitem{webb_high-resolution_2020}
J.J.~Webb and J.~Bovy, \emph{High-resolution simulations of dark matter subhalo
  disruption in a {{Milky}}-{{Way}}-like tidal field},
  \href{https://doi.org/10.1093/mnras/staa2852}{\emph{MNRAS} {\bfseries 499}
  (2020) 116}.

\bibitem{busha_mass_2011}
M.T.~Busha, P.J.~Marshall, R.H.~Wechsler, A.~Klypin and J.~Primack, \emph{{{THE
  MASS DISTRIBUTION AND ASSEMBLY OF THE MILKY WAY FROM THE PROPERTIES OF THE
  MAGELLANIC CLOUDS}}},
  \href{https://doi.org/10.1088/0004-637X/743/1/40}{\emph{ApJ} {\bfseries 743}
  (2011) 40}.

\bibitem{cautun_milky_2014}
M.~Cautun, C.S.~Frenk, R.~{van de Weygaert}, W.A.~Hellwing and B.J.T.~Jones,
  \emph{Milky {{Way}} mass constraints from the {{Galactic}} satellite gap},
  \href{https://doi.org/10.1093/mnras/stu1849}{\emph{MNRAS} {\bfseries 445}
  (2014) 2049}.

\bibitem{bowman_absorption_2018}
J.D.~Bowman, A.E.E.~Rogers, R.A.~Monsalve, T.J.~Mozdzen and N.~Mahesh, \emph{An
  absorption profile centred at 78 megahertz in the sky-averaged spectrum},
  \href{https://doi.org/10.1038/nature25792}{\emph{Nature} {\bfseries 555}
  (2018) 67}.

\bibitem{dayal_reionization_2017}
P.~Dayal, T.R.~Choudhury, V.~Bromm and F.~Pacucci, \emph{Reionization and
  {{Galaxy Formation}} in {{Warm Dark Matter Cosmologies}}},
  \href{https://doi.org/10.3847/1538-4357/836/1/16}{\emph{ApJ} {\bfseries 836}
  (2017) 16}.

\bibitem{chatterjee_ruling_2019}
A.~Chatterjee, P.~Dayal, T.R.~Choudhury and A.~Hutter, \emph{Ruling out
  3~{{keV}} warm dark matter using 21~cm {{EDGES}} data},
  \href{https://doi.org/10.1093/mnras/stz1444}{\emph{MNRAS} {\bfseries 487}
  (2019) 3560}.

\bibitem{boyarsky_21-cm_2019}
A.~Boyarsky, D.~Iakubovskyi, O.~Ruchayskiy, A.~Rudakovskyi and W.~Valkenburg,
  \emph{21-cm observations and warm dark matter models},
  \href{https://doi.org/10.1103/PhysRevD.100.123005}{\emph{Phys. Rev. D}
  {\bfseries 100} (2019) 123005}.

\bibitem{leo_constraining_2020}
M.~Leo, T.~Theuns, C.M.~Baugh, B.~Li and S.~Pascoli, \emph{Constraining
  structure formation using {{EDGES}}},
  \href{https://doi.org/10.1088/1475-7516/2020/04/004}{\emph{JCAP} {\bfseries
  2020} (2020) 004}.

\bibitem{rudakovskyi_can_2020}
A.~Rudakovskyi, D.~Savchenko and M.~Tsizh, \emph{Can {{EDGES}} observation
  favour any dark matter model?},
  \href{https://doi.org/10.1093/mnras/staa2194}{\emph{MNRAS} {\bfseries 497}
  (2020) 3393}.

\bibitem{benitez-llambay_detailed_2020}
A.~{Benitez-Llambay} and C.~Frenk, \emph{The detailed structure and the onset
  of galaxy formation in low-mass gaseous dark matter haloes},
  \href{https://doi.org/10.1093/mnras/staa2698}{\emph{MNRAS} {\bfseries 498}
  (2020) 4887} [\href{https://arxiv.org/abs/2004.06124}{{\ttfamily
  2004.06124}}].

\bibitem{efstathiou_numerical_1985}
G.~Efstathiou, M.~Davis, S.D.M.~White and C.S.~Frenk, \emph{Numerical
  techniques for large cosmological {{N}}-body simulations},
  \href{https://doi.org/10.1086/191003}{\emph{ApJS} {\bfseries 57} (1985) 241}.

\bibitem{power_inner_2003}
C.~Power, J.F.~Navarro, A.~Jenkins, C.S.~Frenk, S.D.M.~White, V.~Springel
  et~al., \emph{The inner structure of {{$\Lambda$CDM}} haloes \textemdash{}
  {{I}}. {{A}} numerical convergence study},
  \href{https://doi.org/10.1046/j.1365-8711.2003.05925.x}{\emph{MNRAS}
  {\bfseries 338} (2003) 14}.

\bibitem{griffen_caterpillar_2016}
B.F.~Griffen, A.P.~Ji, G.A.~Dooley, F.A.~G{\'o}mez, M.~Vogelsberger,
  B.W.~O'Shea et~al., \emph{{{THE CATERPILLAR PROJECT}}: {{A LARGE SUITE OF
  MILKY WAY SIZED HALOS}}},
  \href{https://doi.org/10.3847/0004-637X/818/1/10}{\emph{ApJ} {\bfseries 818}
  (2016) 10}.

\bibitem{simha_modelling_2017}
V.~Simha and S.~Cole, \emph{Modelling galaxy merger time-scales and tidal
  destruction}, \href{https://doi.org/10.1093/mnras/stx1942}{\emph{MNRAS}
  {\bfseries 472} (2017) 1392}
  [\href{https://arxiv.org/abs/1609.09520}{{\ttfamily 1609.09520}}].

\bibitem{springel_simulations_2005}
V.~Springel, S.D.M.~White, A.~Jenkins, C.S.~Frenk, N.~Yoshida, L.~Gao et~al.,
  \emph{Simulations of the formation, evolution and clustering of galaxies and
  quasars}, \href{https://doi.org/10.1038/nature03597}{\emph{Nature} {\bfseries
  435} (2005) 629}.

\bibitem{the_astropy_collaboration_astropy_2013}
{The Astropy Collaboration}, T.P.~Robitaille, E.J.~Tollerud, P.~Greenfield,
  M.~Droettboom, E.~Bray et~al., \emph{Astropy {{A}} community {{Python}}
  package for astronomy},
  \href{https://doi.org/10.1051/0004-6361/201322068}{\emph{A\&A} {\bfseries
  558} (2013) A33}.

\bibitem{the_astropy_collaboration_astropy_2018}
{The Astropy Collaboration}, a.A.M.~{Price-Whelan},
  B.M.~Sip{\textbackslash}Hocz, H.M.~G{\"u}nther, P.L.~Lim, S.M.~Crawford
  et~al., \emph{The {{Astropy Project}}: {{Building}} an {{Open}}-science
  {{Project}} and {{Status}} of the v2.0 {{Core Package}}},
  \href{https://doi.org/10.3847/1538-3881/aabc4f}{\emph{AJ} {\bfseries 156}
  (2018) 123}.

\bibitem{hunter_matplotlib_2007}
J.D.~Hunter, \emph{Matplotlib {{A 2D Graphics Environment}}},
  \href{https://doi.org/10.1109/MCSE.2007.55}{\emph{Comput. Sci. Eng.}
  {\bfseries 9} (2007) 90}.

\bibitem{walt_numpy_2011}
S.~van~der Walt, S.C.~Colbert and G.~Varoquaux, \emph{The {{NumPy Array}}: {{A
  Structure}} for {{Efficient Numerical Computation}}},
  \href{https://doi.org/10.1109/MCSE.2011.37}{\emph{Comput. Sci. Eng.}
  {\bfseries 13} (2011) 22}.

\bibitem{harris_array_2020}
C.R.~Harris, K.J.~Millman, S.J.~{van der Walt}, R.~Gommers, P.~Virtanen,
  D.~Cournapeau et~al., \emph{Array programming with {{NumPy}}},
  \href{https://doi.org/10.1038/s41586-020-2649-2}{\emph{Nature} {\bfseries
  585} (2020) 357}.

\bibitem{rossum_python_1995}
G.~Rossum, \emph{Python reference manual},  Technical {{Report}} {CWI (Centre
  for Mathematics and Computer Science)}, {NLD} (1995).

\bibitem{van_rossum_python_2009}
G.~Van~Rossum and F.L.~Drake, \emph{Python 3 {{Reference Manual}}},
  {CreateSpace}, {Scotts Valley, CA} (2009).

\bibitem{jones_scipy_2011}
E.~Jones, T.~Oliphant and P.~Peterson, \emph{{{SciPy Open}} source scientific
  tools for {{Python}}},  2011.

\bibitem{virtanen_scipy_2020}
P.~Virtanen, R.~Gommers, T.E.~Oliphant, M.~Haberland, T.~Reddy, D.~Cournapeau
  et~al., \emph{{{SciPy}} 1.0: Fundamental algorithms for scientific computing
  in {{Python}}}, \href{https://doi.org/10.1038/s41592-019-0686-2}{\emph{Nature
  Methods} {\bfseries 17} (2020) 261}.

\end{thebibliography}\endgroup

\end{document}